\documentclass[reqno,onecolumn]{amsart}
\usepackage{amsmath,amssymb,xspace}
\usepackage[musical]{sjctensor}

\numberwithin{equation}{section}

\evensidemargin \oddsidemargin \marginparwidth 0.5in \textwidth 6.5in

\newcommand{\gemism}{gravito-electromagnetism\xspace}

\newcommand{\gemic}{gravito-electro\-magnetic\xspace}

\newcommand{\gelec}{gravito-electric\xspace}

\newcommand{\gmag}{gravito-magnetic\xspace}

\newcommand{\emism}{electromagnetism\xspace}

\newcommand{\emic}{electromagnetic\xspace}

\newcommand{\ghat}{\hat{g}}                              
\newcommand{\Ghat}{\hat{G}}                              

\newcommand{\delhat}{\hat{\nabla}}                       

\newcommand{\bfRhat}{\hat{\mathbf{R}}}                   
\newcommand{\bfRdot}{\dot{\mathbf{R}}}                   

\newcommand{\Rdot}{\dot{R}}                              
\newcommand{\Richat}{\widehat{\mathbf{Ric}}}             
\newcommand{\Ric}{{\mathbf{Ric}}}                        
\newcommand{\Ricdot}
  {\overset{\:\:\boldsymbol{.}}{\mathbf{Ric}}}           

\newcommand{\Einhat}{\widehat{\mathbf{Ein}}}             
\newcommand{\Ein}{{\mathbf{Ein}}}                        
\newcommand{\Eindot}
  {\overset{\:\boldsymbol{.}}{\mathbf{Ein}}}             

\newcommand{\cRhat}{\hat{\mathcal{R}}}                   
\newcommand{\cRdot}{\dot{\mathcal{R}}}                   

\newcommand{\cThat}{\hat{\mathcal{T}}}                   
\newcommand{\cTdot}{\dot{\mathcal{T}}}                   
\newcommand{\bscTdot}{\dot{\boldsymbol{\cT}}}            

\newcommand{\Cdot}{{C^\prime}}                           
\newcommand{\cCdot}{{\cC^\prime}}                        

\newcommand{\party}{{\partial_t}}                        

\newcommand{\ttt}[3][]{#1{#2} \tens \ldots \tens #1{#3}}

\begin{document}


\begin{abstract}

A  tensor description of perturbative Einsteinian gravity about an
arbitrary background spacetime is developed. By analogy with the
covariant laws of electromagnetism in spacetime, {\em
gravito-electro\-magnetic} potentials and fields are defined
to  emulate  electromagnetic gauge transformations  under
substitutions belonging to the gauge symmetry group of
perturbative gravitation. These definitions have the advantage
that on a flat background, with the aid of a covariantly constant
timelike vector field, a subset of the linearised gravitational
field equations can be written in a form that is fully analogous
to Maxwell's equations (without awkward factors of 4 and
extraneous tensor fields). It is   shown how the remaining
equations in the perturbed gravitational system restrict the time
dependence of solutions to these equations and thereby prohibit
the existence of propagating vector fields. The induced {\em
\gemic} Lorentz force on a test particle is evaluated in terms of
these fields together with the torque on a small gyroscope. It is
concluded that the analogy of perturbative gravity to Maxwell's
description of electromagnetism can be valuable for
(quasi-)stationary gravitational phenomena but that the analogy
has its limitations.\newline


\end{abstract}



\title{
{\centerline{\bf Gauge Symmetry and Gravito-Electromagnetism}}}

\author[S. J. Clark and R. W. Tucker]{
    S. J. Clark\\
    {$\quad$}\\
    R. W. Tucker\\
    {$\quad$}\\
{\it Department of Physics,\\
    Lancaster University, LA1 4YB, UK\\}
{\tt s.j.clark@lancaster.ac.uk\\
    robin.tucker@lancaster.ac.uk}}

\date{\today}

\maketitle



\tableofcontents


\section{Introduction}

Einstein's theory of gravitation remains a pinnacle in the
evolution of theoretical physics. It offers an overarching
description of phenomena ranging from the familiar behaviour of
Newtonian gravitation to exotic astrophysical events at the
extremes of space and time. Although its modern formulation is in
terms of tensor fields on a manifold with a spacetime structure,
its physical interpretation often
benefits from a choice of observer and an appropriate reference
frame. One of the traditional methods  for extracting information
from Einstein's gravitational field equations is to exploit the
properties of observers in some fiducial background spacetime in
which gravitational physics is either absent or familiar. This
approach has led to various approximation in such backgrounds.
Further reduction is afforded by a ``3+1'' decomposition in which
spacetime tensors are expressed in terms of a field of frames
adapted to some local foliation of spacetime by spacelike
hypersurfaces. More generally frames are afforded by timelike
vector fields, the integral curves of which model ideal observers.
In such a manner it becomes possible to contemplate different
limits in which matter moves slowly  or the gravitational field is
 weak relative to such observers. It is known that Newtonian
gravitation follows from such a limit. Within the framework of
weak gravity non-Newtonian gravitational effects may arise and a
number of experiments have been devised in order to detect such
phenomena as the ``Lense-Thirring effect'' due to
``frame-dragging'' produced by the earth's rotation
\cite{Thirring}, \cite{Thirring1}, \cite{Thirring2},
\cite{Everitt}. The nature of this effect may be detectable by a
small orbiting gyroscope and is analogous to that produced by the
torque on a small magnetic dipole in the presence of the magnetic
field of a fixed magnetic dipole. Indeed the component of weak
gravity (additional to the dominant Newtonian gravitational field)
responsible for this effect is now referred to as the {\em
gravito-magnetic} field. The sensitivity of recently developed
rotation sensors may also be increased to detect post-Newtonian
effects in the future \cite{Packard}.

Several authors \cite{bct}, \cite{harris}, \cite{thorneNZ},
\cite{thorne}, \cite{wald}, \cite{forward}, \cite{Zee},
\cite{ehlers}, \cite{jan}, \cite{mashhoon}, \cite{damour} have
noticed that a subset of the Einstein equations when perturbed
about flat spacetime can be written in a form that looks
remarkably similar to Maxwell's equations with the Newtonian
gravitational field corresponding to the {\em \gelec} field and
mass-currents playing the role of electric currents. Since the
laws of electromagnetism are well studied and understood this
analogy has proved quite fruitful in the {\em \gemic} context
particularly in astrophysical applications. Extended
``astrophysical jet-structures'' are now thought to have their
origin in {\em \gemic} forces. In \cite{NUT} the details of
astrophysical lensing have been explored in terms of parameters in
the NUT metric.

 It is also
amusing to recall that one of the first theories of post-Newtonian
gravitation was formulated by Heaviside in direct analogy with the
then recently formulated theory of electromagnetism by Maxwell. In
the language of the Poincar\'{e}  isometry group it predicted that
gravitation like electromagnetism was mediated by an independent
vector field rather than a second degree tensor field associated
with the metric of spacetime. This difference of course must imply
that any analogy between weak gravity and electromagnetism is
incomplete and most derivations of the {\em \gemic} field
equations take care to point this out. However in our view the
caveats are themselves often incomplete and  a close examination
of various derivations of the {\em \gemic} equations display
significant differences in detail. The difficulty in making
objective comparisons often arises due to the implicit use of a
particular coordinate system (usually adapted to a flat spacetime
background) or a partial gauge fixing. Indeed the question of the
gauge transformations induced on the {\em \gemic} fields from the
underlying gauge covariance of the perturbative Einstein equations
is usually dealt with rather cursorily. This leads one to
contemplate the {\it most useful way} to define the {\em \gemic}
fields in terms of the perturbed components of the spacetime
metric. Different choices are often responsible for the location
of odd factors of 4 that permeate the {\em \gemic} equations
compared with the Maxwell equations. Such choices also have
implications for the form of the induced {\em \gemic} Lorentz
force  (and torque) in terms of the {\em \gemic} fields that enter
into the equation for the motion of a massive point (spinning)
particle. In \cite{mashhoon} \gemic gauge transformations are
discussed from a perspective different from the one presented in
this paper. Here such transformations are explicitly  related to
the gauge symmetry of perturbative gravitation and the definitions
of \gemic fields in turn induce the notions of gravito-magnetic
and gravito-electric coupling strengths.

In this article  a  tensor description of perturbative Einsteinian
gravity about an arbitrary background spacetime is first
constructed. By analogy with the covariant laws of
electromagnetism in spacetime {\em \gemic} potentials and fields
are then defined
to  emulate electromagnetic gauge transformations  under
substitutions belonging to the gauge symmetry group of
perturbative gravitation. These definitions have the advantage
that on a flat background, with the aid of a covariantly constant
timelike vector field, a {\it subset} of the linearised
gravitational field equations can be written in a form that is
{\it fully analogous to Maxwell's equations} (without awkward
factors of 4 and extraneous tensor fields. It is   shown how the
remaining equations in the perturbed gravitational system restrict
the time dependence of solutions to these equations and thereby
prohibit the existence of propagating vector fields. The induced
{\em \gemic} Lorentz force on a test particle is evaluated by
geodesic perturbation  in terms of these fields together with the
torque on a small gyroscope. It is concluded that the analogy of
perturbative gravity to Maxwell's description of electromagnetism
can be valuable for (quasi-)~stationary gravitational phenomena
but that the analogy has its limitations. It has been argued that
such limitations are absent in the approach to  {\em \gemic} based
on properties of the conformal tensor in a spacetime determined by
Einstein's equations. Although this reformulation makes no
reference to perturbative methods the analogy with the structure
of Maxwell's equations is less direct. A tensorial description of
this formulation is given in Appendix \sect{conf}. Throughout this
article the language of tensor fields as multi-linear maps on
vector and co-vector fields is adopted. Co-vector fields are
manipulated using the exterior calculus of differential forms and
Hodge maps. Manifolds are assumed smooth and tensor fields
sufficiently differentiable as required. Notations based on the
tools used are summarised in Appendix \sect{defs} and some
technical computational details are relegated to Appendices
\sect{connex} and \sect{trans}. In order to facilitate comparisons
with other authors certain field redefinitions are discussed in
section \sect{altanalogy} together with the changes induced by
them in the {\em \gemic} field equations. These alternatives are
discussed in the concluding section where the salient features of
this paper are summarised.


\section{The Maxwell Equations}
\label{sec:max}

Maxwell's equations for the electromagnetic field can be written
concisely (in a general spacetime with metric tensor $g$) in terms
of the Faraday 2-form $F$ and current 1-form $\cJ$ on spacetime:
\begin{subequations}
  \begin{equation}
     \label{eq:maxcov1}
     \rmd F = 0 ,
  \end{equation}
  \begin{equation}
     \label{eq:maxcov2}
     \delta F = \cJ .
  \end{equation}
\end{subequations}
Equation \eqn{maxcov1} implies that in a regular domain
\begin{equation}
  F = \rmd A
\end{equation}
for some {\em $1$-form potential} $A$, and  $F$ clearly remains
invariant under {\em electromagnetic gauge transformations} of $A$
which take the form
\begin{equation}
  \label{eq:emgaugetrans}
  A \mapsto A + \rmd \lambda ,
\end{equation}
where $\lambda$ is an arbitrary smooth function on spacetime.
Equation \eqn{maxcov2} implies that the current is conserved
\begin{equation}
  \delta \cJ = 0 .
\end{equation}

In terms of $A$, \eqn{maxcov1} may be written
\begin{equation}
  \Delta A + \rmd \delta A = - \cJ ,
\end{equation}
where $\Delta =-(\delta \rmd +\rmd \delta)$. In the {\em Lorenz
gauge} defined by the condition
\begin{equation}
  \label{eq:transgauge}
  \delta A = 0 ,
\end{equation}
this reduces to the {\em Helmholtz wave equation}
\begin{equation}
  \label{eq:helm}
  \Delta A = - \cJ .
\end{equation}

The equation of motion for a (spinless) test particle with mass
$m$ and charge $q$ moving along a curve $C(\tau)$, parameterized
by proper-time $\tau$, with tangent vector $\Cdot(\tau)$,  in an
arbitrary gravitational field and \emic field $F$  is
\begin{equation}
  \label{eq:qeqnmotion}
  m \del_\Cdot \Cdot + q\, \sharpen{\rmi_\Cdot F} = 0 ,
\end{equation}
where $\nabla$ is the Levi-Civita connection associated with $g$.
The {\em Lorentz force} $\bbF_\rmL$ is defined by
\begin{equation}
  \label{eq:lorentz}
  \bbF_\rmL = - q\, \sharpen{\rmi_\Cdot F} .
\end{equation}

One can decompose $F$ \wrt some unit-normalized
timelike\footnote{Such a vector satisfies $g(V,V)=-1$ with the
choice of signature for $g$ chosen throughout this article.}
vector field $V$ as
\begin{equation}
  \label{eq:fsplit}
  F = \flatten{V} \wedge \bfe + \# \bfb ,
\end{equation}
where $\rmi_V\bfe=0$ and $\rmi_V\#\bfb=0$.  Similarly one may
decompose $\cJ$ as
\begin{equation}
  \label{eq:currentsplit}
  \cJ = \rho \flatten{V} + \bfj
\end{equation}
where $\rmi_V\bfj =0$.  Ideal observers may be associated with the
integral curves of $V$.

The Maxwell field equations \eqn{maxcov1} and \eqn{maxcov2} can
then be written in  (3+1)-form  in terms of $\bfe$, $\bfb$ and $V$
as
\begin{samepage}
\begin{subequations}
  \label{eq:maxall}
  \begin{equation}
     \label{eq:max1}
     \bfd_V \# \bfb + \bsOmega_V \wedge \bfe = 0 ,
  \end{equation}
  \begin{equation}
     \label{eq:max2}
     \bfD_V \bfe + \bscL_V \# \bfb = 0 ,
  \end{equation}
  \begin{equation}
     \label{eq:max4}
     \bfd_V \# \bfe - \bsOmega_V \wedge \bfb = \rho \# 1 ,
  \end{equation}
  \begin{equation}
     \label{eq:max3}
     \bfD_V \bfb - \bscL_V \# \bfe = \# \bfj ,
  \end{equation}
\end{subequations}
\end{samepage}
where the notation is given in the Appendix \sect{defs}. Such
equations exhibit possible ``pseudo-sources'' measured by
non-inertial observers associated with non-parallel vector fields
$V$.

When $V$ is a parallel vector field (e.g. associated with an
observer defining an inertial frame in flat spacetime) these
reduce to the familiar Maxwell equations relating electromagnetic
fields to their sources written in terms of differential forms
\cite{rwt}.  The metric tensor permits one to define the vector
fields: $\bfE = \sharpen{\bfe}$, $\bfB = \sharpen{\bfb}$, $\bfJ =
\sharpen{\bfj}$,  and  (for $V$ parallel) the Maxwell equations
take their more familiar form
\begin{samepage}
\begin{subequations}
  \begin{equation}
    \mathrm{div} \, \bfB = 0 ,
  \end{equation}
  \begin{equation}
    \mathrm{curl} \, \bfE + \frac{\partial {\bfB}}{\partial t} = 0 ,
  \end{equation}
  \begin{equation}
    \mathrm{div} \, \bfE = \rho ,
  \end{equation}
  \begin{equation}
    \mathrm{curl} \, \bfB - \frac{\partial {\bfE}}{\partial t} = \bfJ .
  \end{equation}
\end{subequations}
\end{samepage}
Similarly decomposing the 4-velocity $\Cdot$ \wrt some (unit
timelike parallel) vector field $V$ as
\begin{equation}
  \Cdot = \frac{1}{\sqrt{1 - \bfg(\bfv,\bfv)}} (V + \bfv) ,
\end{equation}
and assuming that the magnitude of $\bfv$ is small ($v^2 =
\bfg(\bfv,\bfv) \ll 1$),
\begin{equation}
  \Cdot = V + \bfv + O(v^2) ,
\end{equation}
the equation of motion for a  charged non-relativistic particle
becomes
\begin{equation}
  \frac{\rmd {\bfv}}{\rmd t} = \frac{q}{m}(\bfE + \bfv \times \bfB) .
\end{equation}


\section{Perturbative Gravitation}

The analogy between  gravitation and electromagnetism to be
discussed follows from a perturbation of the gravitational field
about some ``background'' spacetime geometry. We explore the
constraints on this geometry in order  to execute this analogy to
its fullest extent.


\subsection{The Perturbed Einstein Equations}\indent

A perturbative approach to Einstein's theory of gravitation can be
based on a formal linearization of spacetime geometry about that
determined by some solution of Einstein's equations for the
spacetime metric tensor. A generic perturbation will be identified
with a class of linearizations that belong to the tangent space to
the space of Einstein solutions. Following \cite{stewart} it is
convenient to introduce a 5-dimensional manifold $\cM $ that can
be foliated by hypersurfaces belonging to a one parameter
($\epsilon$) family of spacetimes. The geometry of each leaf
$M_\epsilon$ of the foliation is determined by some second degree
symmetric tensor field $g(\epsilon)$ on $\cM$ where $\epsilon\in
[-1,1]$ that restricts to a Lorentzian metric tensor field on each
leaf. Furthermore it is asserted that all points of one leaf can
be connected to all points on a neighbouring leaf  by a one
parameter diffeomorphism $\varphi_\epsilon : \cM \to \cM$. Thus a
tensor field $T(0)$ on $M_0$ can be related to a tensor field
$T(\epsilon)$ on  $M_\epsilon$ according to
\begin{equation}
  \hat{\varphi}_{-\epsilon} T(\epsilon)= T(0) +
  \epsilon \dot{T}_\cV + O(\epsilon^2)
\end{equation}
where, for some nowhere vanishing vector field $\cV$ that is
nowhere tangent to the spacetime leaves in $\cM$,
\begin{equation}
  \varphi_\epsilon=\exp(\epsilon\cV)
\end{equation}
induces $\hat{\varphi}_\epsilon$ on tensors  and
\begin{equation}
  \dot{T}_\cV = \cL_\cV
  T(\epsilon)\vert_{\epsilon=0} .
\end{equation}
In a local chart with coordinates $\{x^\mu,\epsilon\}$ adapted to
the foliation one may write:
\begin{equation}
  \cV=\frac{\partial}{\partial\epsilon}
  +\xi^\mu(x)\frac{\partial}{\partial x^\mu} .
\end{equation}
The tensor $\dot{T}_\cV$ is said to be a linearization of
$T(\epsilon)$ about $T(0)$ with respect to a choice of the vector
field $\cV$. Since the leaves $M_\epsilon$ are diffeomorphic it is
natural to identify points on distinct leaves that lie on the same
integral curve of $\cV$. Different choices of $\cV$ correspond to
different identifications. If $\dot{T}_{\cV_1}$ and
$\dot{T}_{\cV_2}$ are distinct linearizations then by construction
their difference is generated by a vector field $X$ on $M_0$:
\begin{equation}
  \dot{T}_{\cV_1}\mapsto\dot{T}_{\cV_2}= \dot{T}_{\cV_1}
  + \cL_X T(0) .
\end{equation}
This may be called a gauge transformation of $\dot{T}_{\cV_1}$
induced by $X$. If $\cL_X T(0) =0$ for all $X$ then $T(\epsilon)$
is gauge invariant. In general there is no preferred positive (or
negative) definite metric on $M_0$ that enables one to assign a
natural norm to quantities that are not gauge invariant in this
sense.

Linearized gravity proceeds by writing the covariant physical
metric tensor $g(\epsilon)$ field so that:
\begin{equation}
  \hat{\varphi}_{-\epsilon} g(\epsilon)= g(0)+\epsilon\dot{g}_\cV+
  O(\epsilon^2)
\end{equation}
By common abuse of notation this is simply written;
\begin{equation}
  \label{eq:metric}
  \hat{g}=g+h
\end{equation}
where $h$ is of order $\epsilon$ and higher order terms in
$\epsilon$ are subsequently neglected. This abbreviated notation
hides the choice of $\cV$ in the definition of $h$ and an
alternative choice arises from the {\em gauge transformation}
\begin{equation}
  h \mapsto h + \epsilon\cL_X g
\end{equation}
for any vector field $X$ on the spacetime with {\em background}
metric $g$.  Similarly we introduce the abbreviated notation
$\hat{T} = \hat{\varphi}_{-\epsilon} T(\epsilon)$, $T = T(0)$,
$\dot{T} = \epsilon \dot{T}_\cV$, for some tensor $T$ (thus
$\dot{T}$ is of order $\epsilon$).

The {\em contravariant physical metric tensor field} $\Ghat$ can
be similarly  written in terms of the {\em induced  contravariant
background metric tensor} $G$ and a {\em contravariant
perturbation tensor} $H$
\begin{equation}
  \Ghat = G + H .
\end{equation}
In a smooth local basis of vector fields on  spacetime $\{X_a\}$,
with dual cobasis $ \{e^b\}$ such that $e^b(X_a)=\delta^b{}_a$ for
$a=0,1,2,3$, the induced perturbation tensor  $H$ can be written
as
\begin{equation}
  H = - h_{ab} X^a \tens X^b
\end{equation}
with $g_{ac}g^{cb}=\delta^b{}_a$ (where $g^{ab}$ are the
components of $G$ in the above cobasis) and $X^a = g^{ab} X_b$.
The metric-dual of vector and 1-form fields, and the associated
raising and lowering of indices are defined with respect to the
{\em background} metric tensor. Likewise any operations that
depend on the metric tensor (e.g. the Hodge map) will be defined
\wrt the {\em background} metric unless indicated otherwise.

To derive the perturbed Einstein tensor in terms of $h$  (in some
gauge) write the Levi-Civita connection $\delhat$ with respect to
the physical metric $\ghat$ in terms of the Levi-Civita connection
$\del$ with respect to $g$ so that for any vector field $X$ on
spacetime:
\begin{equation}
  \label{eq:connex}
  \delhat_X = \del_X + \bsgamma_X ,
\end{equation}
Since $\delhat$ and $\del$ are torsion free
\begin{equation}
  \bsgamma_X Y = \bsgamma_Y X ,
\end{equation}
for arbitrary vector fields $X$ and $Y$. It follows that
\begin{equation}
  \bsgamma_X f = 0 ,
\end{equation}
for  any function $f$  on spacetime. It is convenient to define a
tensor $\gamma$ by
\begin{equation}
  \gamma(X,Y,\alpha) = \alpha(\bsgamma_X Y) ,
\end{equation}
 for any 1-form $\alpha$,
then
\begin{equation}
  \bsgamma_X Y = \gamma(X,Y,-)
\end{equation}
and
\begin{equation}
  \bsgamma_X \alpha = -\gamma(X,-,\alpha) .
\end{equation}

Since $\delhat$ and $\del$ are compatible \wrt $\ghat$ and $g$
respectively  $\gamma$ can be written in terms of $h$ as
\begin{equation}
  \label{eq:pertconnex}
  \gamma(X,Y,\alpha) = \frac{1}{2}\{(\del_X
  h)(Y,\sharpen{\alpha})+(\del_Y
  h)(X,\sharpen{\alpha})-(\del_{\sharpen{\alpha}} h)(X,Y)\} ,
\end{equation}
for arbitrary $X$, $Y$ and $\alpha$.

%
%

The curvature operators $\bfRhat$ and $\bfR$ for the connections
$\delhat$ and $\del$ respectively are defined in the usual manner
as
\begin{subequations}
  \begin{equation}
     \bfRhat_{X\:Y} = [\delhat_X,\delhat_Y] - \delhat_{[X,Y]} ,
  \end{equation}
  \begin{equation}
     \bfR_{X\:Y} = [\del_X,\del_Y] - \del_{[X,Y]} .
  \end{equation}
\end{subequations}
Thus
\begin{equation}
  \bfRhat_{X\:Y} = \bfR_{X\:Y} + \bfRdot_{X\:Y} ,
\end{equation}
and it  follows that the {\em perturbed curvature operator}
$\bfRdot$ is given by
\begin{equation}
  \bfRdot_{X\:Y} = \del_X \bsgamma_Y - \del_Y \bsgamma_X + \bsgamma_X
  \del_Y - \bsgamma_Y \del_X - \bsgamma_{[X,Y]} .
\end{equation}
When acting on any vector field $Z$ this simplifies to
\begin{equation}
  \bfRdot_{X\:Y} Z = (\del_X \gamma)(Y,Z,-) - (\del_Y \gamma)(X,Z,-) ,
\end{equation}
and  the {\em perturbed curvature tensor} $\Rdot$ is defined by
\begin{equation}
  \Rdot(X,Y,Z,\alpha) = \alpha(\bfRdot_{X\:Y} Z) = (\del_X
  \gamma)(Y,Z,\alpha) - (\del_Y \gamma)(X,Z,\alpha) .
\end{equation}

The {\em perturbed Ricci tensor} $\Ricdot$ follows by contraction
\begin{equation}
  \Ricdot(X,Y) = \Rdot(X_a,X,Y,e^a) .
\end{equation}
It is useful to introduce the {\em trace-reverse map} $\mu$ on
covariant degree 2 tensors so that
\begin{equation}
  \mu(T) = T - \tfrac{1}{2}\Tr(T)\, g
\end{equation}
where $ \Tr(T)=T(X_a,X^a)$ in any basis. The trace-reverse of $h$
is denoted by $\psi$:
\begin{equation}
  \psi =\mu(h) .
\end{equation}
After some calculation $\Ricdot$ can be written in terms of $\psi$
and the Laplacian operator $\Lap=\div \del$
as
\begin{equation}
  \Ricdot = - \frac{1}{2} \mu(\Lap \psi) + \Sym \del \div \psi -
  \cC_\psi + \cS_\psi ,
\end{equation}
where
\begin{subequations}
  \begin{equation}
     \cC_\psi (X,Y) = R(X_a,X,Y,\psi^a) ,
  \end{equation}

  \begin{equation} \cS_\psi(X,Y) =
     \frac{1}{2}(\Ric(\sharpen{\psi}_X,Y) + \Ric(\sharpen{\psi}_Y,X))
      \end{equation}
\end{subequations}
and  the convenient notation
\begin{subequations}
  \begin{equation}
     \psi_X = \psi(X,-) ,
  \end{equation}
  \begin{equation}
     \psi^a = \psi_{X^a} = \psi(X^a,-)
  \end{equation}
\end{subequations}
is used.

Note that $\cC_\psi$ and $\cS_\psi$ have the same trace, namely
\begin{equation}
  \Tr(\cC_\psi) = \Tr(\cS_\psi) = \Ric(\sharpen{\psi}_a,X^a) .
\end{equation}

%
%

The {\em perturbed curvature scalar} $\cRdot$ follows as
\begin{equation}
  \cRdot = \Tr(\Ricdot - \cS_\psi + \frac{1}{2} \cR \psi) ,
\end{equation}
and with $\Einhat = \Richat - \tfrac{1}{2} \ghat \cRhat = \Ein +
\Eindot$ the {\em perturbed Einstein tensor} $\Eindot$ is
\begin{equation}
  \Eindot = \Ricdot - \frac{1}{2} \cR h - \frac{1}{2} \cRdot g ,
\end{equation}
or
\begin{equation}
  \Eindot = - \frac{1}{2} \Lap \psi + \mu(\Sym \del \div \psi) -
  \mu(\cC_\psi) + \cS_\psi - \frac{1}{2} \cR \psi
\end{equation} in terms of $\psi$.
%
%

One may now express the {\em perturbed Einstein equation} in terms
of $\psi$ by writing the {\em physical stress energy-momentum
tensor} $\cThat$ as
\begin{equation}
  \cThat = \cT + \cTdot ,
\end{equation}
where the {\em background stress energy-momentum tensor} $\cT$
acts as a source for the background metric via the {\em background
Einstein equation}
\begin{equation}
  \label{eq:backein}
  \Ein = \kappa \cT ,
\end{equation}
and $\kappa = 8 \pi G$ in units where $c = 1$. The {\em perturbed
Einstein equation}
\begin{equation}
  \label{eq:eindot}
  \Eindot = \kappa \cTdot ,
\end{equation}
becomes
\begin{equation}
  - \frac{1}{2} \Lap \psi + \mu(\Sym \del \div \psi) - \mu(\cC_\psi)
  + \cS_\psi - \frac{1}{2} \cR \psi = \kappa \cTdot .
\end{equation}
In the next section this equation is simplified by exploiting a
{\em gauge symmetry} of the Einstein equations.



\subsection{Gauge Transformations and the Transverse Gauge Condition}\indent

A {\em gauge transform} of $h$ has been  defined as a substitution
of the form
\begin{equation}
  h \mapsto h + \cL_V g ,
  \label{eq:basic}
\end{equation}
where $\cL_V$ is the Lie derivative with respect to some vector
field $V$ that maintains $h$ perturbative with respect to $g$.
This substitution is used to determine the induced gauge
transformation of tensors defined in terms of $h$ (keeping the
background geometry fixed). Thus the induced gauge transformation
of $\bsgamma_X$ follows as
\begin{equation}
  \bsgamma_X \mapsto \bsgamma_X + \cD_X (V) ,
\end{equation}
where, for any vector field  $X$  the tensor derivation $\cD_X
(V)$ is defined by
\begin{equation}
  \cD_X (V) = [\cL_V,\del_X] - \del_{\cL_V X} .
\end{equation}
The operator $ \cD_X (V)$ provides a useful tool when performing
calculations involving both Lie derivatives and covariant
derivatives (\cite{rwt}).

%
%

It follows that the perturbed curvature operator transforms as
\begin{equation}
  \bfRdot_{X\:Y} \mapsto \bfRdot_{X\:Y} + \cR_{X\:Y} (V) ,
\end{equation}
where the operator $\cR_{X\:Y}(V)$ is defined by
\begin{equation}
  \cR_{X \: Y} (V) = [\cL_V,\bfR_{X \: Y}] - \bfR_{\cL_V X \:Y}
  - \bfR_{X \: \cL_V Y} .
\end{equation}
A contraction shows that the perturbed curvature tensor transforms
as
\begin{equation}
  \Rdot \mapsto \Rdot + \cL_V R ,
\end{equation}
from which one deduces
\begin{equation}
  \Ricdot \mapsto \Ricdot + \cL_V \Ric
\end{equation}
and
\begin{equation}
  \cRdot \mapsto \cRdot + \cL_V \cR .
\end{equation}
It follows that the perturbed Einstein tensor exhibits the induced
gauge transformation
\begin{equation}
  \Eindot \mapsto \Eindot + \cL_V \Ein .
\end{equation}
This behaviour of $\Eindot$ dictates the behaviour of any
(phenomenological) stress energy-momentum  tensor  under an
induced gauge transformation in order to maintain the gauge
covariance of \eqn{eindot}.

%
%

It can be similarly shown that if $\psi \mapsto
\bar{\psi}=\psi+\mu(\cL_V g)$ the divergence of $\psi$ transforms
as
\begin{equation}
  \div \psi \mapsto \div \bar{\psi}=
\div \psi + \flatten{\Lap V} + \Ric(V,-) .
\end{equation}
Thus, if for some $\psi$, one  chooses $V$ to satisfy the
differential equation
\begin{equation}
  \flatten{\Lap V} + \Ric(V,-) = - \div \psi ,
\end{equation}
then this imposes the gauge condition
\begin{equation}
  \label{eq:gravtransgauge}
  \div \bar{\psi} = 0 ,
\end{equation}
referred to as the {\em transverse gauge}. Further gauge
transformations, generated by additional vector fields  $W$
satisfying the linear differential equation
\begin{equation}
  \flatten{\Lap W} + \Ric(W,-) = 0 ,
\end{equation}
maintain $\bar{\psi}$ divergenceless.

%
%

In the transverse gauge the perturbed Einstein equation
immediately simplifies to
\begin{equation} - \frac{1}{2} \Lap \psi - \mu(\cC_\psi) + \cS_\psi
   - \frac{1}{2} \cR \psi = \kappa \cTdot .
\end{equation}
It should be noted that this equation holds in {\em any}
(background) metric that satisfies the (background) Einstein
equation \eqn{backein}.

In the case where the background metric is Ricci-flat (or has a
cosmological constant), the perturbed Einstein equation simplifies
to
\begin{equation}
  - \frac{1}{2} \Lap \psi - \cC_\psi = \kappa \cTdot
\end{equation}
which may be compared with the component form \cite{mtw} and the
abstract index form \cite{wald}.

If the background metric  is flat this simplifies further to
\begin{equation}
  \label{eq:flatein}
  \Lap \psi = - 2 \kappa \cTdot .
\end{equation}
This equation is analogous to the Helmholtz equation for the
electromagnetic potential 1-form in the electromagnetic Lorenz
gauge and leads to the prediction of propagating gravitational
perturbations.


\section{Gravito-Electromagnetism}\label{sec:lingem}

The perturbed Einstein equation about a {\em flat} background
spacetime metric has a  form that is similar to the Helmholtz
equation for the electromagnetic potential. However the perturbed
gravitational potential $\psi$ is a symmetric type (2,0) tensor
field, whereas the electromagnetic potential $A$ is a type (1,0)
tensor field (a 1-form). The description of Maxwell's equations in
terms of electric and magnetic fields suggests that $\psi$ be
decomposed relative to some fiducial vector field and split  into
a frame dependent 1-form (analogous to the 1-form potential in
\emism) plus an extra non-Maxwellian tensor field.
We motivate this definition by inducing a gauge transformation
described above on this 1-form and comparing the result with the
structure of the  electromagnetic gauge transformation (Section
\sect{max}) of the 1-form potential.
This will also show that restrictions must be placed on the
fiducial vector field if a close analogy with \emism is to hold.


\subsection{Gravito-Electromagnetic Gauge Transformations}\indent
\label{sec:gauge}

The trace-reversed perturbation $\psi$ can be written, relative to
some unit-normalized (to zero order in $\epsilon$) timelike vector
field $\xi$, ($ g(\xi,\xi)= -1 + O(\epsilon)$) and general
background metric tensor $g$ as
\begin{equation}
  \label{eq:psisplit} \psi = \phi_\xi g - \psi_\xi \tens
  \flatten{\xi} - \flatten{\xi} \tens \psi_\xi - \bsSigma_\xi .
\end{equation}
\begin{subequations}
The tensor $\psi$ has been split into a 1-form part
  \begin{equation}
     \psi_\xi = \psi(\xi,-) ,
  \end{equation}
with $\xi$ component
  \begin{equation}
     \phi_\xi = \psi_\xi(\xi) = \psi(\xi,\xi) ,
  \end{equation}
and a spacelike tensor part $\bsSigma_\xi$ satisfying
  \begin{equation}
     \bsSigma_\xi(\xi,-) = 0 .
  \end{equation}
\end{subequations}
Since
\begin{equation}
  \label{eq:psitrans}
  \psi \mapsto \psi + \mu(\cL_V g)
\end{equation}
under a gauge transformation  $h \mapsto h + \cL_V g $ generated
by $V,$ a contraction with  $\xi$ gives
\begin{equation}
  \psi_\xi \mapsto \psi_\xi + \rmd (g(\xi,V)) + \rmi_V \rmd
  \flatten{\xi} - \div V \flatten{\xi} + \flatten{\cL_\xi V} .
\end{equation}
With a special choice for $V$ of the form
\begin{equation}
  \label{eq:gemtrans}
  V = - \lambda \xi ,
\end{equation}
where $\lambda$ is any smooth function of order $\epsilon$ on
spacetime,  the induced transformation on the 1-form $\psi_\xi$
becomes
\begin{equation}
  \psi_\xi \mapsto \psi_\xi + \rmd \lambda + \lambda (\Theta_\xi
  \flatten{\xi} - \flatten{\bscA}_\xi) ,
\end{equation}
where $\Theta_\xi$ is the expansion of $\xi$ and $\bscA_\xi$ is
its acceleration (Appendix \sect{defs}).  If
 $\lambda (\Theta_\xi
  \flatten{\xi} - \flatten{\bscA}_\xi)$ is of higher order in $\epsilon$
   than $\lambda$ the last
  term
   can be neglected and the gauge transformation of $\psi_\xi$ simplifies to
\begin{equation}
   \label{eq:psixitrans}
   \psi_\xi \mapsto \psi_\xi + \rmd \lambda .
\end{equation}
This suggests that  $\psi_\xi$  be interpreted as the analogue of
the 1-form potential $A$ in \emism and  henceforth $\psi_\xi$ will
be referred to as the {\em \gemic $1$-form potential}.

%
%

Similarly, projecting \eqn{psitrans}  with    $\bsPi_\xi$
(Appendix \sect{defs}) one finds
\begin{equation}
  \bsSigma_\xi \mapsto \bsSigma_\xi - \bsPi_\xi \cL_V g + 2(\div
  V + g(\xi,\del_\xi V)) \bfg_\xi .
\end{equation}
Hence with $V$ as in \eqn{gemtrans}, $\bsSigma_\xi$ transforms as
\begin{equation}
  \bsSigma_\xi \mapsto \bsSigma_\xi + 2 \lambda (\bssigma_\xi -
  \tfrac{2}{3} \Theta_\xi \bfg_\xi) ,
\end{equation}
where $\bssigma_\xi$ is the shear of $\xi$ (Appendix \sect{defs}).
If $ \lambda (\bssigma_\xi - \tfrac{2}{3} \Theta_\xi \bfg_\xi)$ is
of higher order in $\epsilon$ than $\bsSigma_\xi $ the last term
is negligible and $\bsSigma_\xi$ is gauge invariant under such
transformations.

%
%

The class of gauge transformations generated by $- \lambda \xi$
such that $\psi_\xi$ transforms as in \eqn{psixitrans} with
$\bsSigma_\xi$ invariant will be said to contain {\em \gemic gauge
transformations}. Clearly if $\xi$ is parallel (i.e. $\del \xi =
0$), or parallel to  zero order in $\epsilon$ (since $\lambda$ is
of order $\epsilon$), such transformations are guaranteed to be
\gemic gauge transformations.  For a given $h$ one may therefore
define a class of vector fields $\{\xi\}$, members of which are
equivalent if they can be used to generate \gemic gauge
transformations. Such fields will be said to define {\em \gemic
frames of reference}.
Not all background spacetimes will permit the existence of such
vector fields (e.g. the black-hole Schwarzschild spacetime),
however they are guaranteed to exist in Minkowski spacetime (and
interestingly also in the Einstein static universe).
In the next section these vector fields will be used to define a
class of gauge equivalent {\em gravito-electromagnetic} fields.


\subsection{The Gravito-Electromagnetic Field Equations}\indent
\label{sec:gemfieldeq}

On a flat background the linearized Einstein equation can be
written as
\begin{equation}
  \label{eq:linein}
  \Lap \psi = - 2 \kappa \cTdot ,
\end{equation}
where the transverse gauge condition
\begin{equation}
  \label{eq:transgc}
  \div \psi = 0 ,
\end{equation}
has been imposed.   Contracting \eqn{linein} with a
unit-normalized timelike vector field $\xi$ that can be used to
define \gemic gauge transformations gives
\begin{equation}
   \label{eq:vecwaveeqn}
   \Delta \psi_\xi = 2 \kappa \cJ_\xi ,
\end{equation}
where the {\em mass-current} $\cJ_\xi$ is defined by
\begin{equation}
  \label{eq:source}
  \cJ_\xi = - \cTdot(\xi,-) .
\end{equation}

Contracting \eqn{vecwaveeqn} on $\xi$ gives
\begin{equation}
  \label{eq:scalwaveeqn}
  \Delta \phi_\xi = - 2 \kappa \rho_\xi ,
\end{equation}
where the {\em mass-density} $\rho_\xi$ is defined by
\begin{equation}
  \label{eq:density}
  \rho_\xi = - \cJ_\xi(\xi) = \cTdot(\xi,\xi).
\end{equation}
Similarly acting on \eqn{linein} with the projection operator
$\bsPi_\xi$ and using \eqn{psisplit}
\begin{equation}
  \Delta \phi_\xi \bfg_\xi - \Lap \bsSigma_\xi = - 2 \kappa
  \bscTdot_\xi.
\end{equation}
With the aid of \eqn{scalwaveeqn} this becomes
\begin{equation}
  \Lap \bsSigma_\xi = 2 \kappa (\bscTdot_\xi - \rho_\xi \bfg_\xi) ,
  \label{eq:lapsigmaeqn}
\end{equation}
where $\bscTdot_\xi$ is the spacelike (\wrt $\xi$), part of the
stress  energy-momentum tensor
\begin{equation}
  \label{eq:slsource}
  \bscTdot_\xi = \bsPi_\xi \cTdot .
\end{equation}

The \gemic analogue of the Faraday 2-form is now defined as
\begin{equation}
  \cF_\xi = \rmd \psi_\xi ,
\end{equation}
hence
  \begin{equation}
    \label{eq:gemmaxcov1}
    \rmd \cF_\xi = 0 .
  \end{equation}


The tensor $\cF_\xi$ is invariant under transformations in the
class of \gemic gauge transformation defined above. The  {\em
\gelec field} $\bscE_\xi$, and the {\em \gmag field} $\bscB_\xi$
follow  from a (3+1)-split (\wrt $\xi$):
\begin{equation}
  \label{eq:eb31def}
  \cF_\xi = \flatten{\xi} \wedge \bscE_\xi + \# \bscB_\xi .
\end{equation}
With $\psi_\xi$ written in terms  of its timelike and spacelike
parts:
\begin{equation}
  \label{eq:psixisplit}
  \psi_\xi = - \phi_\xi \flatten{\xi} + \bsPsi_\xi ,
\end{equation}
where
$\rmi_\xi \bsPsi_\xi=0$, one can  write the \gelec field as
\begin{subequations}
  \begin{equation}
     \label{eq:gefield}
     \bscE_\xi = \bfd_\xi \phi_\xi - \bscL_\xi \bsPsi_\xi  ,
  \end{equation}
  and the \gmag field as
  \begin{equation}
     \label{eq:gmfield}
     \bscB_\xi = \# \bfd_\xi \bsPsi_\xi
  \end{equation}
\end{subequations}
in terms of the \gemic potential.
 The mass-current can be similarly split into its
timelike and spacelike parts  with respect to $\xi$ as
\begin{equation}
  \cJ_\xi = \rho_\xi \flatten{\xi} + \bfJ_\xi ,
\end{equation}
where
 $\rmi_\xi\bfJ_\xi=0$ .

The transverse gauge condition in terms of $\xi$, $\psi_\xi$,
$\phi_\xi$ and $\bsSigma_\xi$   is:
\begin{equation}
  \div \psi = \delta \psi_\xi \flatten{\xi} + \rmd \phi_\xi -
  \del_\xi \psi_\xi - \div \bsSigma_\xi - \del_{\sharpen{\psi}_\xi}
  \flatten{\xi} - \Theta_\xi \psi_\xi = 0 .
\end{equation}
For a \gemic frame $\xi$ ($\nabla \xi=0$ to order $\epsilon$)
 this reduces to
\begin{equation}
  \div \psi = \delta \psi_\xi \flatten{\xi} + \rmd \phi_\xi -
  \del_\xi \psi_\xi - \div \bsSigma_\xi = 0.
\end{equation}
Using \eqn{psixisplit}, \eqn{dell} and \eqn{dsplit} it follows
that $\rmd \phi_\xi - \del_\xi \psi_\xi = \bfd_\xi \phi_\xi -
\bscL_\xi \bsPsi_\xi = \bscE_\xi$, hence
\begin{equation}
  \label{eq:transsplit}
  \div \psi = (\delta \psi_\xi \flatten{\xi}) + ( \bscE_\xi - \div
  \bsSigma_\xi) = 0 .
\end{equation}
The two bracketed terms are orthogonal 
 so
the transverse gauge condition is equivalent to the two equations
\begin{subequations}
   \begin{equation}
      \label{eq:firstgc}
      \delta \psi_\xi = 0 ,
   \end{equation}
   \begin{equation}
      \label{eq:secondgc}
      \bscE_\xi = \div \bsSigma_\xi .
   \end{equation}
\end{subequations}
The first condition is analogous to the Lorenz gauge in \emism,
while the second condition has no \emic analogue.  The
consequences of this second condition are explored below.

The equation \eqn{firstgc} implies (Appendix \sect{defs}) that
\eqn{vecwaveeqn} takes the Maxwell-like covariant form
  \begin{equation}
    \label{eq:gemmaxcov2}
    \delta \cF_\xi = - 2 \kappa \cJ_\xi .
  \end{equation}

The perturbative part of the  stress energy-momentum tensor may be
expressed in terms of the mass-current using \eqn{source},
\eqn{density} and \eqn{slsource} as
\begin{equation}
  \cTdot = \rho_\xi \flatten{\xi} \tens \flatten{\xi} - \cJ_\xi
  \tens \flatten{\xi} - \flatten{\xi} \tens \cJ_\xi + \bscTdot_\xi ,
\end{equation}
and since the background source is assumed zero ($\cT = 0$) the
divergence condition becomes
\begin{equation}
  \div \cTdot = 0
\end{equation}
hence
\begin{equation}
  (\delta \cJ_\xi \flatten{\xi}) - (\del_\xi \bfJ_\xi - \div
  \bscTdot_\xi) = 0 .
\end{equation}
Since the two bracketed terms are orthogonal one has:
\begin{subequations}
   \begin{equation}
      \label{eq:firstcl}
      \delta \cJ_\xi = 0 ,
   \end{equation}
   \begin{equation}
      \label{eq:secondcl}
      \del_\xi \bfJ_\xi = \div \bscTdot_\xi .
   \end{equation}
\end{subequations}
with the first condition expressing the conservation of
mass-current in the background geometry.

The field equation \eqn{gemmaxcov2} and the closure of $\cF_\xi$
\eqn{gemmaxcov1} can now be written in terms of these
decompositions:
\begin{samepage}
  \begin{subequations}
    \label{eq:gemmaxall}
    \begin{equation}
       \label{eq:gemmax1}
       \bfd_\xi \# \bscB_\xi = 0 ,
    \end{equation}
    \begin{equation}
       \label{eq:gemmax2}
       \bfd_\xi \bscE_\xi + \bscL_\xi \# \bscB_\xi = 0 ,
    \end{equation}
    \begin{equation}
       \label{eq:gemmax4}
       \bfd_\xi \# \bscE_\xi = - 2 \kappa \rho_\xi \# 1 ,
    \end{equation}
    \begin{equation}
       \label{eq:gemmax3}
       \bfd_\xi \bscB_\xi - \bscL_\xi \# \bscE_\xi = - 2 \kappa \# \bfJ_\xi .
    \end{equation}
  \end{subequations}
\end{samepage}


Although the equations \eqn{gemmaxall} involving the gravitational
field have the mathematical structure of Maxwell's equations for
electromagnetism the two theories are not isomorphic since the
gravitational fields must additionally satisfy \eqn{secondgc} in
the transverse gauge. Not all solutions of Maxwell's equations
translate to perturbative gravitational fields that are compatible
with this condition for a given $\bsSigma_\xi$. There does  exist
however a class of compatible solutions. Such solutions $\psi$ are
characterized by the existence of a gauge in which
\begin{subequations}
\begin{equation}
   \div \psi = 0 ,
\end{equation}
\begin{equation}
   \label{eq:gemlimit}
   \bsPi_\xi \psi = 0 ,
\end{equation}
\end{subequations}
in some  \gemic frame $\xi$ (Section \sect{gauge}). In this case
such solutions will be said to belong to the {\em \gemic limit}.
It follows from equation \eqn{psisplit} that
\begin{equation}
  \bsSigma_\xi = \phi_\xi \bfg_\xi .
  \label{eq:gemlimiteqn}
\end{equation}
Such a $\xi$ will not be unique. If $\zeta = \xi + \bfv$, such
that $g(\zeta,\zeta) = -1$ and $v^2 = g(\bfv,\bfv) \ll 1$, then
$\bsPi_\zeta \psi$ is of  order $v\epsilon$. Hence if $v \lesssim
\epsilon$, $\zeta$ also  defines the \gemic limit. However  the
fields ($\psi_\zeta$, $\bsSigma_\zeta$, $\bscE_\zeta$,
$\bscB_\zeta$ ) defined \wrt $\zeta$ are identical (within this
approximation) to the corresponding fields defined \wrt $\xi$
(Appendix \sect{trans}).

 The condition \eqn{gemlimit}  and the field equation
\eqn{linein}
 require
\begin{equation}
  \label{eq:sourcelimit}
  \bscTdot_\xi = 0 .
\end{equation}
Equation \eqn{gemlimit} implies $\div \bsSigma_\xi = \bfd_\xi
\phi_\xi$, hence in this limit condition  \eqn{secondgc} becomes
\begin{equation}
  \label{eq:egradphi}
  \bscE_\xi = \bfd_\xi \phi_\xi ,
\end{equation}
or with \eqn{gefield}
\begin{equation}
  \bscL_\xi \bsPsi_\xi = 0 ,
\end{equation}
and using \eqn{firstgc} it follows that
\begin{equation}
  \xi^2 \phi_\xi = 0 .
\end{equation}
Thus  $\bsPsi_\xi$ is time independent and it follows from
\eqn{gmfield} that the gravito-magnetic field is also time
independent, $\bscL_\xi \bscB_\xi = 0$ . In this limit the field
equations reduce to
\begin{samepage}
  \begin{subequations}
    \label{eq:gemmaxallgc}
    \begin{equation}
       \label{eq:gemmax1gc}
       \bfd_\xi \# \bscB_\xi = 0 ,
    \end{equation}
    \begin{equation}
       \label{eq:gemmax2gc}
       \bfd_\xi \bscE_\xi = 0 ,
    \end{equation}
    \begin{equation}
       \label{eq:gemmax4gc}
       \bfd_\xi \# \bscE_\xi = - 2 \kappa \rho_\xi \# 1 ,
    \end{equation}
    \begin{equation}
       \label{eq:gemmax3gc}
       \bfd_\xi \bscB_\xi - \bscL_\xi \# \bscE_\xi = - 2 \kappa \# \bfJ_\xi ,
    \end{equation}
  \end{subequations}
\end{samepage}
along with the conditions \eqn{firstgc}, \eqn{secondgc} which can
be written as
\begin{subequations}
  \begin{equation}
  \label{eq:firstgcsplit}
  \bfd_\xi \# \bsPsi_\xi - \xi \phi_\xi = 0 ,
\end{equation}
\begin{equation}
  \label{eq:secondgcsplit}
  \bscL_\xi \bsPsi_\xi = 0 .
\end{equation}
\end{subequations}
respectively. Similarly it follows from \eqn{firstcl},
\eqn{secondcl} and \eqn{sourcelimit} that
\begin{subequations}
\begin{equation}
  \label{eq:firstclsplit}
  \bfd_\xi \# \bfJ_\xi + \xi \rho_\xi = 0 ,
\end{equation}
\begin{equation}
  \label{eq:secondclsplit}
  \bscL_\xi \bfJ_\xi = 0 ,
\end{equation}
\end{subequations}

With $\xi=\frac{\partial}{\partial t}$ and the vector fields
$\bbE_\xi = \sharpen{\bscE}_\xi$, $\bbB_\xi =
\sharpen{\bscB}_\xi$, $\bbJ_\xi = \sharpen{\bfJ}_\xi$, the field
equations \eqn{gemmaxallgc} can be written in more familiar
notation as
\begin{subequations}
  \begin{equation}
    \mathrm{div} \, \bbB_\xi = 0 ,
  \end{equation}
  \begin{equation}
    \mathrm{curl} \, \bbE_\xi = 0 ,
  \end{equation}
  \begin{equation}
    \mathrm{div} \, \bbE_\xi = - 2 \kappa \rho_\xi ,
  \end{equation}
  \begin{equation}
    \mathrm{curl} \, \bbB_\xi - \frac{\partial {\bbE}_\xi}{\partial t} =
    - 2 \kappa \bbJ_\xi.
  \end{equation}
\end{subequations}
Similarly, with $\bbA_\xi = \sharpen{\bsPsi}_\xi$, the gauge
conditions \eqn{firstgcsplit},
 \eqn{secondgcsplit} become
\begin{subequations}
  \begin{equation}
  \mathrm{div} \bbA_\xi - \frac{\partial {\phi}_\xi}{\partial t} = 0 ,
\end{equation}
\begin{equation}
  \frac{\partial {\bbA}_\xi}{\partial t} = 0
\end{equation}
\end{subequations}
and the conservation equations \eqn{firstclsplit},
\eqn{secondclsplit} become
\begin{subequations}
\begin{equation}
  \mathrm{div} \bbJ_\xi + \frac{\partial {\rho}_\xi}{\partial t} = 0 ,
\end{equation}
\begin{equation}
  \frac{\partial {\bbJ}_\xi}{\partial t} = 0 .
\end{equation}
\end{subequations}
Thus stationary electric and magnetic field configurations and
their sources have direct analogues in the theory of perturbative
gravitation.

The {\em Newtonian limit} is defined  as the \gemic limit
supplemented by the condition
\begin{equation}
   \bsPsi_\xi = 0 .
\end{equation}
 From \eqn{vecwaveeqn} this
implies
\begin{equation}
   \bfJ_\xi = 0 .
\end{equation}
The {\em Newtonian potential} $\Phi_\xi$ is identified as
\begin{equation}
  \label{eq:newtpot}
  \Phi_\xi = - \tfrac{1}{4} \phi_\xi .
\end{equation}From \eqn{firstgcsplit} it must be time-independent
\begin{equation}
  \xi \Phi_\xi = 0 ,
\end{equation}
and equation \eqn{scalwaveeqn} becomes
\begin{equation}
  \label{eq:newtfe}
  \Delta \Phi_\xi = 4 \pi G \rho_\xi ,
\end{equation}
which is just the field equation for Newtonian gravitation.

In the following section the above framework is illustrated by
calculating the \gemic fields arising as perturbations on the
asymptotic gravitational field of a rotating source.


\subsection{The \gemic Field of a Rotating Source}\indent


The metric tensor at large distances from a compact rotating body,
with Newtonian gravitational mass $M$ and angular momentum $J$,
may be approximated (for $r \gg 2GM$ in units where $c=1$) by
\begin{equation}
\begin{aligned}
  \ghat = & \; - \left(1 - \frac{2GM}{r} \right) \rmd t \tens \rmd t
  + \left(1 + \frac{2GM}{r}\right) (\rmd r \tens \rmd r
  + r^2 \rmd \theta \tens \rmd \theta
  + r^2 \sin^2 \theta \rmd \phi \tens \rmd \phi) \\
  & + \frac{2GJ}{r^2} \sin \theta ( r \sin \theta \rmd \phi \tens \rmd t
   + r \sin \theta \rmd t \tens \rmd \phi)  .
\end{aligned}
\end{equation}
In the cobasis $\{e^0 = \rmd t, e^1 = \rmd r, e^2 = r \rmd \theta,
e^3 = r \sin \theta \rmd \phi\}$ the metric tensor can be
rewritten as
\begin{equation}
\begin{aligned}
  \ghat = & \; - \left(1 - \frac{R_\rmS}{r} \right) e^0 \tens e^0
  + \left(1 + \frac{R_\rmS}{r}\right) (e^1 \tens e^1
  + e^2 \tens e^2 + e^3 \tens e^3) \\
  & + \frac{R_\rmS R_\rmK}{r^2} \sin \theta (e^0 \tens e^3 + e^3 \tens e^0)
\end{aligned}
\end{equation}
in terms of the two length-scales $R_\rmS = 2 G M$ and $R_\rmK =
\frac{J}{M}$.

Defining the dimensionless coordinates $R$ and $T$ by
\begin{subequations}
  \begin{equation}
    r = \frac{R_\rmS}{\epsilon} R ,
  \end{equation}
  \begin{equation}
    t = \frac{R_\rmS}{\epsilon} T ,
  \end{equation}
\end{subequations}
and the dimensionless cobasis $\underline{e}^a$ by
\begin{equation}
  e^a = \frac{R_\rmS}{\epsilon} \underline{e}^a ,
\end{equation}
the dimensionless physical metric tensor $\underline{\ghat}$
defined by
\begin{equation}
  \ghat = \left(\frac{R_\rmS}{\epsilon}\right)^2 \underline{\ghat} ,
\end{equation}
can be rewritten as
\begin{equation}
\begin{aligned}
  \label{eq:dimphysmet}
  \underline{\ghat} = & \; - \left(1 - \frac{\epsilon}{R} \right) \underline{e}^0 \tens \underline{e}^0
  + \left(1 + \frac{\epsilon}{R}\right) (\underline{e}^1 \tens \underline{e}^1
  + \underline{e}^2 \tens \underline{e}^2 + \underline{e}^3 \tens \underline{e}^3) \\
  & + \frac{\epsilon}{R^2} \Lambda \sin \theta (\underline{e}^0 \tens \underline{e}^3 + \underline{e}^3 \tens \underline{e}^0)  ,
\end{aligned}
\end{equation}
where the constant $\Lambda$ is
\begin{equation}
  \Lambda = \epsilon \frac{R_\rmK}{R_\rmS} .
\end{equation}
For the last term in \eqn{dimphysmet} not to be second order in
$\epsilon$ (and hence negligible), $\frac{R_\rmS}{R_\rmK}$ must be
of order $\epsilon$.

With the dimensionless Minkowski metric tensor
\begin{equation}
  \underline{g} = - \underline{e}^0 \tens \underline{e}^0 + \underline{e}^1 \tens \underline{e}^1 + \underline{e}^2 \tens \underline{e}^2 + \underline{e}^3 \tens \underline{e}^3
\end{equation}
the perturbation $\underline{h}$ about $\underline{g}$ is then
\begin{equation}
  \underline{h} = \frac{\epsilon}{R} (\underline{e}^0 \tens \underline{e}^0
      + \underline{e}^1 \tens \underline{e}^1 + \underline{e}^2 \tens \underline{e}^2
      + \underline{e}^3 \tens \underline{e}^3)
      + \frac{\epsilon}{R^2} \Lambda \sin \theta ( \underline{e}^3 \tens \underline{e}^0 + \underline{e}^0 \tens \underline{e}^3) ,
\end{equation}
and taking the trace-reverse yields
\begin{equation}
  \underline{\psi} = 2\frac{\epsilon}{R} \underline{e}^0 \tens \underline{e}^0
       + \frac{\epsilon}{R^2} \Lambda \sin \theta ( \underline{e}^3 \tens \underline{e}^0 +
        \underline{e}^0 \tens \underline{e}^3) .
\end{equation}

Using the dimensionless basis $\{\underline{X}_a\}$ dual to
$\{\underline{e}^a\}$ given by $\{\underline{X}_0 = \partial_T,
\underline{X}_1 = \partial_R, \underline{X}_2 = \frac{1}{R}
\partial_\theta, \underline{X}_3 = \frac{1}{R \sin \theta}
\partial_\phi\}$, we use $\underline{X}_0$ to define the
dimensionless \gemic 1-form potential
$\underline{\psi}_{\underline{X}_0}$ as
\begin{equation}
 \label{eq:dimlesspsi}
  \underline{\psi}_{\underline{X}_0} = 2
  \frac{\epsilon}{R} \underline{e}^0 + \frac{\epsilon}{R^2} \Lambda \sin \theta \underline{e}^3 .
\end{equation}
In terms of the dimensionless gravito-Faraday 2-form
$\underline{\cF}_{\underline{X}_0}$ defined by
\begin{equation}
  \underline{\cF}_{\underline{X}_0} = \rmd \underline{\psi}_{\underline{X}_0} ,
\end{equation}
 the dimensionless \gelec and \gmag fields,
$\underline{\bscE}_{\underline{X}_0}$ and
$\underline{\bscB}_{\underline{X}_0}$ follow by writing
\begin{equation}
  \underline{\cF}_{\underline{X}_0} = \underline{X}^{\underline{\flat}}_0 \wedge \underline{\bscE}_{\underline{X}_0}
   + \underline{\#}\underline{\bscB}_{\underline{X}_0} .
\end{equation}
Hence from \eqn{dimlesspsi}
\begin{equation}
  \underline{\bscE}_{\underline{X}_0} = - 2\frac{\epsilon}{R^2} \underline{e}^1 ,
\end{equation}
and (with  $\underline{\#} 1 = \underline{e}^1 \wedge
\underline{e}^2 \wedge \underline{e}^3$) the dimensionless \gmag
field is
\begin{equation}
  \underline{\bscB}_{\underline{X}_0} =
  \frac{\epsilon}{R^3} \Lambda (\sin \theta \underline{e}^2 + 2 \cos \theta \underline{e}^1) .
\end{equation}

In terms of $Z = R \cos \theta$, $\underline{\sharp}$ the
metric-dual map associated with $\underline{g}$ and
\begin{equation}
  \bsLambda = \Lambda \partial_Z ,
\end{equation}
the gravito-electric and -magnetic vector fields defined by
$\underline{\bbE}_{\underline{X}_0}=
\underline{\bscE}^{\underline{\sharp}}_{\underline{X}_0}$ and
$\underline{\bbB}_{\underline{X}_0}=
\underline{\bscB}^{\underline{\sharp}}_{\underline{X}_0}$ become
\begin{subequations}
  \begin{equation}
    \label{elec}
    \underline{\bbE}_{\underline{X}_0} = - 2\frac{\epsilon}{R^2} \underline{X}_1 ,
  \end{equation}
  \begin{equation}
    \label{mag}
    \underline{\bbB}_{\underline{X}_0}
     = - \frac{\epsilon}{R^3} (\bsLambda -
     3 \underline{g}(\bsLambda,  \underline{X}_1)
     \underline{X}_1).
  \end{equation}
\end{subequations}
These fields may be compared with those derived in \cite{thorneNZ}
and \cite{thorne}. In the next two sections the motion of a test
particle in a \gemic field is discussed. For such a particle with
``spin'' the motion is expected to follow from the weak field
limit of equations presented in \cite{Dixon}. To facilitate a
comparison with treatments that ignore Mathisson-Papapetrou type
coupling to spacetime curvature the discussion is restricted to
the separate pre-geodesic motion of a spinless test particle and
the gyroscopic precession that follows from a parallel spin vector
along such a motion.


\section{Motion of a Test Particle in a Weak Gravitational Field}\indent

The history of an electrically neutral, spinless test particle in
spacetime is modeled by a future-pointing timelike curve $C(\tau)$
that satisfies
\begin{equation}
  \label{eq:pregeodesic}
  \delhat_\Cdot \Cdot - \frac{1}{2} \frac{\Cdot(\ghat(\Cdot,\Cdot))}{\ghat(\Cdot,\Cdot)} \Cdot = 0 ,
\end{equation}
for some general parameter $\tau$.  To compare the following
\gemic equations with  analogous equations from electromagnetism
{\em in Minkowski spacetime} it is most convenient to parameterize
the curve so that
\begin{equation}
  g(\Cdot,\Cdot) = -1 .
\end{equation}

Using \eqn{metric} and \eqn{connex}, to first order in $\epsilon$
equation \eqn{pregeodesic} is
\begin{equation}
   \label{eq:xixi}
     \del_\Cdot \Cdot + \gamma(\Cdot,\Cdot,-) - \frac{1}{2} \Cdot(h(\Cdot,\Cdot)) \Cdot = 0
     .
\end{equation}

Rewriting $\gamma(\Cdot,\Cdot,-)$ in terms of $\cF_\Cdot$ and
$\bsSigma_\Cdot$ (see Appendix \sect{connex}), \eqn{xixi} becomes
\begin{equation}
   \label{eq:eqnmotion}
   \del_\Cdot \Cdot + \sharpen{\rmi_\Cdot
   \cF_\Cdot + \tfrac{1}{4} \bfd_\Cdot \Tr(\bsSigma_\Cdot)} = 0
\end{equation}
and, for a test particle with mass $m$, the perturbed {\em
gravitational force} $\bbF_\rmG$, is then
\begin{equation}
   \label{eq:gemlorentz}
   \bbF_\rmG = - m \sharpen{\rmi_\Cdot
   \cF_\Cdot + \tfrac{1}{4} \bfd_\Cdot \Tr(\bsSigma_\Cdot)} .
\end{equation}
The first term on the right is analogous to the Lorentz force in
\emism, however the second term is non-Maxwellian. Since
$\bbF_\rmG$ depends only on $\cF_\Cdot$ and $\bsSigma_\Cdot$ it is
invariant under \gemic gauge transformations (Section
\sect{gauge}).

%
%

To examine the above equations in the \gemic limit  assume that
some vector $\party$ defines a \gemic frame, and reparameterize
$C(\tau)$ in terms of $t$, so that $\cC(t) = C(\tau)$.  Writing
$\cCdot(t)$ as
\begin{equation}
  \label{eq:cdotv}
  \cCdot = \party + \bfv ,
\end{equation}
where $g(\bfv,\party)=0$, it then follows that
\begin{equation}
  g(\cCdot,\cCdot) = - (1-v^2) ,
\end{equation}
where $v^2 = g(\bfv,\bfv)$.

Writing $\Cdot = \frac{1}{\sqrt{1-v^2}} \cCdot$, and introducing
the assumption
 that the speed $v$ of the particle is non-relativistic, so that $v \ll 1$
 (hence terms smaller than $\epsilon v$ will be neglected), \eqn{eqnmotion}
  can be written in terms of $\cCdot$ as
\begin{equation}
   \label{eq:eqnmotion2}
   \del_\cCdot \cCdot + v \party v \, \cCdot + \sharpen{\rmi_\cCdot
   \cF_\cCdot + \tfrac{1}{4} \bfd_\cCdot \Tr(\bsSigma_\cCdot)} = 0 .
\end{equation}

Expressing $\cF_\cCdot$ and $\bsSigma_\cCdot$ in terms of
$\psi_{\party}$
 (see Appendix \sect{trans} with $\zeta=\cCdot$ and $\xi=\party$) so that
\begin{subequations}
\begin{equation}
  \cF_\cCdot = \cF_\party + \bfd_\party (\psi_\party (\bfv)) \wedge \rmd t ,
\end{equation}
\begin{equation}
  \bsSigma_\cCdot = \{ \phi_\party + 2 \psi_\party (\bfv) \} \bfg_\party - \psi_\party \tens \flatten{\bfv} + \flatten{\bfv} \tens \psi_\party ,
\end{equation}
\end{subequations}
it follows that
\begin{subequations}
\begin{equation}
  \rmi_\cCdot \cF_\cCdot = - \bscE_\party - \# (\flatten{\bfv} \wedge \bscB_\party ) - \bfd_\party (\psi_\party (\bfv))  + \bscE_\party (\bfv) \rmd t ,
\end{equation}
\begin{equation}
  \bfd_\cCdot \Tr(\bsSigma_\cCdot) = 3 \bscE_\party + 4 \bfd_\party (\psi_\party (\bfv)) + 3 \party \phi_\party \bfv - 3 \bscE_\party (\bfv) \rmd t ,
\end{equation}
\end{subequations}
where \eqn{egradphi} has been used.

Using \eqn{cdotv} the spatial part of the equation of motion
\eqn{eqnmotion2} becomes
\begin{equation}
   \frac{\rmd {\bfv}}{\rmd t} =  \tfrac{1}{4} \sharpen{\bscE}_\party
   + \# \sharpen{\flatten{\bfv} \wedge \bscB_\party} - \frac{3}{4} \party \phi_\party \bfv ,
\end{equation}
and if $\phi_\party$ is time-independent this reduces to
\begin{equation}
   \frac{\rmd {\bfv}}{\rmd t} =  \tfrac{1}{4} \sharpen{\bscE}_\party
   + \# \sharpen{\flatten{\bfv} \wedge \bscB_\party} .
\end{equation}
or in vector notation
\begin{equation}
  \frac{\rmd {\bfv}}{\rmd t} = \tfrac{1}{4}\bbE_\party + \bfv \times \bbB_\party ,
\end{equation}
where $\bbE_\party = \sharpen{\bscE}_\party$ and $\bbB_\party =
\sharpen{\bscB}_\party$.

In terms of these fields the perturbed gravitational force
$\bbF_\rmG$ is then
\begin{equation}
  \label{eq:pop}
  \bbF_\rmG =  m \left( \tfrac{1}{4}\bbE_\party + \bfv \times \bbB_\party \right) .
\end{equation}
This is similar in structure to the Lorentz force law of \emism,
except for the factor of $\tfrac{1}{4}$ multiplying the \gelec
term (and the fact that the $\bbE_\party$ and $\bbB_\party$ fields
couple universally to inertial mass).
 Note that rescaling $\bbE_\party$ to
remove the factor $\tfrac{1}{4}$ would introduce a factor into the
gravito-Maxwell equations above.

If one continues to the Newtonian limit then $\bbB_\party = 0$
and, using \eqn{egradphi}, \eqn{newtpot} to express $\bbE_\party$
in terms of the Newtonian potential $\Phi_\party$ (which will
automatically be time-independent by the gauge condition $\div
\psi = 0$), \eqn{pop} reduces to
\begin{equation}
  \bbF_\rmG = - m \, \mathrm{grad} \, \Phi_\party ,
\end{equation}
which is just Newton's force of gravity in terms of a
gravitational potential $\Phi_\party$ satisfying \eqn{newtfe}.


\section{Precession of a Small Gyroscope in a Weak  Gravitational Field}\indent

One may model the relativistic  spin of a freely falling gyroscope
by a unit spacelike vector field  $\bfS$  along a future-pointing
timelike curve $C(\tau)$ that satisfies \eqn{pregeodesic}, such
that $\bfS$ solves
\begin{equation}
  \label{eq:delZS}
  \delhat_\Cdot \bfS = 0 ,
\end{equation}
with
\begin{subequations}
  \begin{equation}
     \ghat(\bfS,\bfS) = 1 ,
  \end{equation}
  \begin{equation}
     \ghat(\bfS,\Cdot) = 0 ,
  \end{equation}
\end{subequations}
in terms of the physical metric tensor.  A perturbative analysis
can be given in terms of  the vector $\bfs$ defined along
$C(\tau)$ by
\begin{equation}
  \label{eq:newspin}
  \bfs = (1 + \tfrac{1}{2} h(\bfS,\bfS)) \bfS - h(\bfS,\Cdot) \Cdot ,
\end{equation}
so that (to first order in $\epsilon$)
\begin{subequations}
  \begin{equation}
     \label{eq:spinnorm}
     g(\bfs,\bfs) = 1 ,
  \end{equation}
  \begin{equation}
     g(\bfs,\Cdot) = 0 ,
  \end{equation}
\end{subequations}
in terms of the background metric $g$.

Equation \eqn{newspin} can be inverted and to first order in
$\epsilon$
\begin{equation}
  \bfS = (1 - \tfrac{1}{2} h(\bfs,\bfs)) \bfs + h(\bfs,\Cdot) \Cdot .
\end{equation}
Using the perturbed connection, \eqn{delZS} can be written in
terms of $\bfs$ to first order in $\epsilon$ as
\begin{equation}
  \del_\Cdot \bfs - \frac{1}{2} \Cdot (h(\bfs,\bfs)) \bfs + \Cdot( h(\bfs,\Cdot)) \Cdot + \gamma(\Cdot,\bfs,-) = 0 .
\end{equation}
Rewriting $\gamma(\Cdot,\bfs,-)$ in terms of $\cF_\cCdot$ and
$\bsSigma_\cCdot$ (see Appendix \sect{connex}), this becomes
\begin{equation}
  \label{eq:spineqn}
  \del_\Cdot^{\rmF} \bfs +
  \frac{1}{2}\sharpen{\bsPi_\Cdot \rmi_\bfs
  \cF_\Cdot - \bsPi_\bfs \del_\Cdot \{ \bsSigma_\Cdot (\bfs,-) \} } = 0,
\end{equation}
where for any vector field $X$, the {\em Fermi-Walker connection}
$\del^\rmF$ is defined on $C$ by
\begin{equation}
  \del_\Cdot^{\rmF} X = \del_\Cdot X + g(\Cdot,X) \bscA_\Cdot -
  g(\bscA_\Cdot,X) \Cdot ,
\end{equation}
with $\bscA_\Cdot = \del_\Cdot \Cdot$.  From \eqn{spineqn}
  the effective gravitational torque $\bbT_\rmG$ on the gyroscope is
\begin{equation}
  \label{eq:torque}
  \bbT_\rmG = - \frac{1}{2}\sharpen{\bsPi_\Cdot \rmi_\bfs
  \cF_\Cdot - \bsPi_\bfs \del_\Cdot \{ \bsSigma_\Cdot (\bfs,-) \} }  .
\end{equation}

The vector $\bfs$ is said to be non-rotating along the path $C$
when $\bbT_\rmG=0$. As with the point particle, if this curve is
parameterized in terms of $t$ so that $\cCdot(t)$ is given by
\eqn{cdotv} (and dropping terms smaller than $\epsilon v$),
\eqn{spineqn} becomes
\begin{equation}
  \label{eq:spineq2}
  \del_\cCdot^{\rmF} \bfs +
  \frac{1}{2}\sharpen{\bsPi_\cCdot \rmi_\bfs
  \cF_\cCdot - \bsPi_\bfs \del_\cCdot \{ \bsSigma_\cCdot (\bfs,-) \} } = 0 .
\end{equation}
The vector $\bfs$  can be written in terms of its spacelike
component $\bssigma$ as
\begin{equation}
  \bfs = g(\bssigma,\bfv) \partial_t + \bssigma,
\end{equation}
which follows from \eqn{spinnorm}.  Imposing the \gemic limit in
the $\party$ frame, and writing $\cF_\cCdot$ and $\bsSigma_\cCdot$
in terms of $\bbE_\party$, $\bbB_\party$ and $\phi_\party$ (see
Appendix \sect{trans}), the spacelike component of \eqn{spineq2}
then becomes
\begin{equation}
  \label{eq:spineq3}
  \frac{\rmd {\bssigma}}{\rmd t} = \frac{1}{2} \left( \bssigma \times \bbB_\party + (\bssigma \cdot \bfv) \bbE_\party  -  \frac{1}{2} (\bssigma \cdot \bbE_\party) \bfv + (\cCdot \phi_\party) \bssigma  \right) .
\end{equation}

Since the length $\sqrt{\bssigma \cdot \bssigma}$ of the  vector
$\bssigma$ is dependent on $t$ (to the appropriate order),
introduce the constant length  gyroscopic spin vector $\bbS$ by
\begin{equation}
  \bbS = (1 - \tfrac{1}{2} \phi_\party) \bssigma - \tfrac{1}{2} (\bfv \cdot \bssigma) \bfv .
\end{equation}
  In terms of $\bbS$, \eqn{spineq3} can be written
\begin{equation}
 \label{eq:gyroeqn}
  \frac{\rmd {\bbS}}{\rmd t} =
   \frac{1}{2} \bbS \times \left(\bbB_\party - \frac{3}{4} \bfv \times \bbE_\party \right)
\end{equation}
which may be compared with the similar equation found in
\cite{harris}.

To this approximation the precession rate
$\frac{1}{2}\left(\frac{3}{4} \bfv \times\bbE_\party - \bbB_\party
\right)$ of $\bbS$ is independent of the gyroscopic spin and could
in principle be used to detect $\bbB_\party$.



\section{Field Redefinitions}
\label{sec:altanalogy}

In order to facilitate a comparison with alternative formulations
of weak gravity using the  \gemic analogy it is worthwhile to
effect certain field redfinitions. These are motivated by looking
at the analogues  of the  transformations \eqn{basic} with
\eqn{gemtrans}.

Let $h_\xi = h(\xi,-)$ and $\xi$ be a \gemic frame, then under a
\gemic gauge transformation $h_\xi$ transforms as
\begin{equation}
  h_\xi \mapsto h_\xi + \rmd \lambda - (\xi \lambda) \flatten{\xi} ,
\end{equation}
which is not quite of the same form as \eqn{emgaugetrans}.
However, defining
\begin{subequations}
  \begin{equation}
    \Phi_h = - \tfrac{1}{2} h(\xi,\xi) ,
  \end{equation}
and
  \begin{equation}
    \bfA_h = \bsPi_\xi \sharpen{h}_\xi ,
  \end{equation}
\end{subequations}
it follows, writing $\xi = \party$, that
\begin{subequations}
  \label{eq:altgemtrans}
  \begin{equation}
    \Phi_h \mapsto \Phi_h - \party \lambda ,
  \end{equation}
  \begin{equation}
    \bfA_h \mapsto \bfA_h + \mathrm{grad} \, \lambda
  \end{equation}
\end{subequations}
which is indeed analogous to the (3+1)-decomposed form of an \emic
gauge transformation.

Similarly $\psi_\xi$ transforms as in \eqn{psixitrans} and
defining
\begin{subequations}
  \begin{equation}
   \Phi_\psi = - \phi_\xi = - \psi(\xi,\xi) ,
  \end{equation}
  \begin{equation}
    \bfA_\psi = \bsPi_\xi \sharpen{\psi}_\xi ,
  \end{equation}
\end{subequations}
the (3+1)-decomposed form of the gauge transformation
\eqn{psixitrans} becomes
\begin{subequations}
  \label{eq:decompgemtrans}
  \begin{equation}
    \Phi_\psi \mapsto \Phi_\psi - \party \lambda ,
  \end{equation}
  \begin{equation}
    \bfA_\psi \mapsto \bfA_\psi + \mathrm{grad} \, \lambda .
  \end{equation}
\end{subequations}

Since  \eqn{altgemtrans}, \eqn{decompgemtrans} are analagous to
\emic gauge transformations this suggests that  a Maxwell type
system of field equations may be   constructed using $\Phi_h$ and
$\bfA_h$ instead of $\Phi_\psi$ and $\bfA_\psi$.

>From \eqn{psisplit} and the above definitions of $\Phi_h$,
$\Phi_\psi$, $\bfA_h$ and $\bfA_\psi$ it follows that
\begin{subequations}
  \begin{equation}
    \Phi_\psi = \Phi_h - \tfrac{1}{4} \Tr(\bsSigma) ,
  \end{equation}
  \begin{equation}
    \bfA_\psi = \bfA_h ,
  \end{equation}
\end{subequations}
where $\bsSigma = \bsSigma_\xi$.

The Lorenz gauge condition in \eqn{firstgc} written in terms of
$\Phi_\psi$ and $\bfA_\psi$ becomes
\begin{subequations}
  \begin{equation}
    \mathrm{div} \, \bfA_\psi + \party \Phi_\psi = 0 ,
    \label{eq:divaeqn}
  \end{equation}
which has the same form as in \emism for all $\bsSigma$. In terms
of $\Phi_h$ and $\bfA_h$ this becomes
  \begin{equation}
    \label{eq:hlorenz}
    \mathrm{div} \, \bfA_h + \party \Phi_h = \tfrac{1}{4} \Tr(\bsSigma) ,
  \end{equation}
\end{subequations}
which is unlike the \emic Lorenz gauge condition when  the
right-hand side is non-zero.

The \gelec and -magnetic fields  $\bfE_\psi = \bbE_\xi$ and
$\bfB_\psi = \bbB_\xi$ discussed earlier are related to
$\Phi_\psi$ and $\bfA_\psi$ by
\begin{subequations}
  \begin{equation}
    \bfE_\psi = - \mathrm{grad} \, \Phi_\psi - \party \bfA_\psi ,
  \end{equation}
  \begin{equation}
    \bfB_\psi = \mathrm{curl} \, \bfA_\psi .
  \end{equation}
\end{subequations}

Alternative \gemic fields $\bfE_h$ and $\bfB_h$ can be defined in
terms of $\Phi_h$ and $\bfA_h$ as
\begin{subequations}
  \begin{equation}
    \bfE_h = - \mathrm{grad} \, \Phi_h - \party \bfA_h ,
  \end{equation}
  \begin{equation}
    \bfB_h = \mathrm{curl} \, \bfA_h .
  \end{equation}
\end{subequations}
These are related to $\bfE_\psi$ and $\bfB_\psi$ by
\begin{subequations}
  \begin{equation}
    \label{eq:ehepsirel}
    \bfE_\psi = \bfE_h + \tfrac{1}{4} \mathrm{grad} \Tr(\bsSigma) ,
  \end{equation}
  \begin{equation}
    \label{eq:bhbpsirel}
    \bfB_\psi = \bfB_h .
  \end{equation}
\end{subequations}
In terms of $\bfE_\psi$ and $\bsSigma$ the second gauge condition
\eqn{secondgc} can be written as
\begin{subequations}
  \begin{equation}
    \label{eq:nonmaxgcpsi}
    \bfE_\psi = \sharpen{\div \bsSigma} ,
  \end{equation}
or using $\bfE_h$ and $\bsSigma$ as
  \begin{equation}
    \label{eq:nonmaxgch}
    \bfE_h = \sharpen{\div \bsSigma} - \tfrac{1}{4} \mathrm{grad} \Tr(\bsSigma) .
  \end{equation}
\end{subequations}
In the $\bfE_\psi$ and $\bfB_\psi$ notation , with  $ \kappa= 8\pi
G$,  $\rho = \rho_\xi$ and $\bfJ = \bbJ_\xi$ the field equations
\eqn{gemmaxall} become
\begin{subequations}
  \begin{equation}
    \mathrm{div} \, \bfB_\psi = 0 ,
  \end{equation}
  \begin{equation}
    \mathrm{curl} \, \bfE_\psi + \party \bfB_\psi = 0 ,
  \end{equation}
  \begin{equation}
    \mathrm{div} \, \bfE_\psi = -16 \pi G \rho ,
  \end{equation}
  \begin{equation}
    \mathrm{curl} \, \bfB_\psi - \party \bfE_\psi = -16 \pi G \bfJ
  \end{equation}
\end{subequations}
for all $\bsSigma$. As stressed above these equations together
with \eqn{lapsigmaeqn}, \eqn{divaeqn}, \eqn{nonmaxgcpsi} summarise
linearised  Einsteinnian gravitation.

Alternatively using \eqn{ehepsirel} and \eqn{bhbpsirel}  the above
  take the form
\begin{subequations}
  \label{eq:hgemfield}
  \begin{equation}
    \mathrm{div} \, \bfB_h = 0 ,
  \end{equation}
  \begin{equation}
    \mathrm{curl} \, \bfE_h + \party \bfB_h = 0 ,
  \end{equation}
  \begin{equation}
    \label{eq:gemsigcoup1}
    \mathrm{div} \, \bfE_h = -16 \pi G \rho - \tfrac{1}{4} \mathrm{div} \, \mathrm{grad} \, \Tr(\bsSigma) ,
  \end{equation}
  \begin{equation}
    \label{eq:gemsigcoup2}
    \mathrm{curl} \, \bfB_h - \party \bfE_h = -16 \pi G \bfJ + \tfrac{1}{4} \party \, \mathrm{grad} \,
    \Tr(\bsSigma)
  \end{equation}
\end{subequations}
in terms of $\bfE_h$ and $\bfB_h$ and $\bsSigma$.
 In the \gemic
limit, $\Tr(\bsSigma) = - 12 \Phi_h$ so  $\Phi_\psi = 4 \Phi_h$
and $\bfE_\psi = 4 \bfE_h$. Then the  Lorenz gauge condition
\eqn{hlorenz} becomes
\begin{equation}
  \mathrm{div} \, \bfA_h + 4 \party \Phi_h = 0 ,
\end{equation}
and (since $\party \bfB_h = 0$ in the \gemic limit) the field
equations \eqn{hgemfield} simplify to
\begin{subequations}
  \begin{equation}
    \mathrm{div} \, \bfB_h = 0 ,
  \end{equation}
  \begin{equation}
    \mathrm{curl} \, \bfE_h = 0 ,
  \end{equation}
  \begin{equation}
    \mathrm{div} \, \bfE_h = -4 \pi G \rho ,
  \end{equation}
  \begin{equation}
    \mathrm{curl} \, \bfB_h - 4 \party \bfE_h = -16 \pi G \bfJ
  \end{equation}
\end{subequations}
which may be compared with (4.44) and the formulations given in
 \cite{thorneNZ},  \cite{harris}.


\section{Discussion}
\label{sec:discuss}

In this article the analogy between  Maxwell's equations for the
electromagnetic field,  \eqn{maxcov1}, \eqn{maxcov2} and the
Einstein equations for weak gravitational fields in the transverse
gauge, \eqn{gemmaxcov1}, \eqn{gemmaxcov2} has been made in terms
of tensor fields. While the former are valid in an arbitrary
Lorentzian spacetime the latter have been developed in terms of
perturbations about a flat spacetime background. A comparison has
 been made between the general equation describing the motion of
electrically charged point particles  \eqn{qeqnmotion} and the
motion of massive point particles in a weak gravitational field
\eqn{eqnmotion}. Equations have also been developed \eqn{gyroeqn}
 describing the motion of a freely falling small gyroscope in terms
of  \gemic fields.

In general it is asserted that any analogy between
electromagnetism and weak gravity is closest in a restricted class
of reference frames related by suitable non-relativistic
transformations and for stationary physical field configurations
in such frames. In addition to  \gemic fields the general
equations of weak gravity involve
 a second degree  symmetric tensor field $\bsSigma_\xi$ which has no
electromagnetic analogue. The \gemic fields defined in
\ref{sec:gemfieldeq} are coupled to $\bsSigma_\xi$  via
\eqn{secondgc} and this field produces non-Maxwellian terms in the
weak gravitational force and torque equations, \eqn{gemlorentz}
and \eqn{torque} respectively.

In \emism  the Maxwell fields $A$, $F$ and $\cJ$ are  defined
independent of any  frame of reference. The latter    is only
required to define electric and magnetic fields and their sources
in terms of electric charge and current density  \eqn{fsplit},
\eqn{currentsplit}. In \gemism the analogous fields  $\psi_\xi$,
$\cF_\xi$ and $\cJ_\xi$ are manifestly frame-dependent.

The definition of these fields has been motivated by their
behaviour under a class of gauge transformations belonging to the
gauge symmetry of the weak Einstein equations.  Unlike \emism,
$\cF_\xi$ is not gauge invariant under these transformations in
general. However  a subset of these transformations does exist for
which  $\cF_\xi$ remains invariant.  These have been called \gemic
gauge transformations by analogy with the gauge symmetry of
Maxwell's equations. Unlike electromagnetic interactions with
electrically charged particles,  weak gravitational interactions
are not mediated by complex representations of these symmetries.
In terms of the \gemic fields a subset of the linearised Einstein
equations take a remarkable form that is isomorphic to Maxwell's
\emic field equations. No limits or further approximations are
required to establish this correspondence. The explicit appearance
of the $\bsSigma_\xi$ tensor is confined to the remaining
equations in the linearised system. This is a primary distinction
of the approach adopted here compared with previous derivations of
the \gemic field equations. To exploit this reformulation and link
 physical field configurations with solutions to Maxwell's
 equations further conditions must be imposed on the linearised
 system.

  Conditions have been found that enable
a useful analogy between weak gravitation and electromagnetism to
be established. In the \gemic limit $\bsSigma_\xi$ depends only on
the a \gemic potential   \eqn{gemlimiteqn}. Consequently, such weak
gravitational fields can be described in terms of $\psi_\xi$ (or
alternatively $\bscE_\xi$ and $\bscB_\xi$).

However as stressed above, in order to obtain  Maxwell-like
equations the transverse gauge condition \eqn{transgc} is imposed
on $\psi$. Condition \eqn{transsplit} induces an equivalent
condition \eqn{firstgc} on $\psi_\xi$. (By contrast the
electromagnetic  gauge condition \eqn{transgauge} is one of many
that may be imposed  on $A$.) The condition on $\psi$ also imposes
the restriction \eqn{secondgc} which in the \gemic limit implies
that $\psi_\xi$ has a restricted time dependence. Physical \gemism
consequently shares more in common  with electromagneto-statics
than electromagnetism. Although both the  Maxwell equations
\eqn{maxall}  and the equivalent \gemic equations \eqn{gemmaxall}
hold in any inertial frame (in Minkowski spacetime),
 in order to remain within the \gemic limit only  a class of \gemic frames
of reference is permitted (Appendix \sect{trans}).

By perturbing the equation of a physical timelike  geodesic the
relativistic equation of motion for a massive point particle in
the weak gravitational field can be cast into a form containing a
\gemic Lorentz force \eqn{gemlorentz} and an additional
non-Maxwellian term.
 It is worth pointing out that in
the context of the \gemic fields defined in this paper the
derivation of this equation of motion does not rely on the speed
of the particle and the \gemic Lorentz-like force takes its
natural form.
 In the \gemic and non-relativistic limit the particle acceleration is
 then determined by a non-relativisti
Lorentz-like force \eqn{pop}  containing an additional  factor of
$\frac{1}{4}$ multiplying the \gelec field.

%
%

When working in the \gemic limit the field redefinitions presented
in section \sect{altanalogy} permit a comparison with the work of
Thorne in \cite{thorneNZ} with the notation $\Phi_h = \Phi$,
$\bfA_h = \bsgamma$, $\bfE_h = \bfg$, and $\bfB_h = \bfH$, and
with a current of the form $\bfJ = \rho \bfv$. The work of
\cite{harris}, \cite{mashhoon}, \cite{bct}, \cite{thorne} and
others may be related to that of \cite{thorneNZ} either by trivial
field redefinitions or changes in metric signature.
 The analogy between
electromagnetism and weak gravity developed by Wald \cite{wald} is
similar to the one presented here. However he does not discuss how
the \gemic 1-form-potential behaves under gauge transformations
nor how the restricted time dependence arises from the transverse
gauge condition.

This article  offers an alternative description of weak field
gravitation in the language of \gemic fields. We feel that it
clarifies a number of issues concerning various other analogies
between the equations of post-Newtonian gravity according to
Einstein and Maxwell's description of electromagnetism. In the
absence of a \gemic limit formulations based on $\Phi_h$ and
$\bfA_h$ give rise to  gauge conditions \eqn{hlorenz},
\eqn{nonmaxgch} and field equations \eqn{gemsigcoup1},
\eqn{gemsigcoup2} thereby  exposing couplings between the \gemic
potentials and fields and  the tensor field $\bsSigma_\xi$. In the
approach adopted here such couplings are relegated to  the gauge
condition \eqn{nonmaxgcpsi}. Many analogies coalesce in the \gemic
limit modulo a re-shuffling of numerical factors that cannot be
scaled away entirely.  Although the mathematical analogy between
weak gravity and the full system of Maxwells equations for
electromagnetism in terms of covariant tensor fields on flat
spacetime can be made close, the existence of gravitational gauge
conditions limits the physical\footnote{Fields derived from a
potential $\phi_{\partial_t}$ that vary linearly with $t$ are
deemed unphysical here.} analogy to stationary phenomena. Despite
this limitation the interpretation of weak gravity in terms of
\gemic fields offers a fertile avenue of exploration for phenomena
associated with the detection of the stationary gravito-magnetic
field. The methods presented here are also applicable in principle
to certain non-flat backgrounds and to weak field descriptions of
non-Einsteinian gravitation (in which other geometrical fields may
compete with metric-induced gravity) at the post-Newtonian level.
These issues will be discussed elsewhere.

\section{Acknowledgement}
SJC is grateful to PPARC and  RWT to  BAE-Systems  for support for
this research. The authors are also grateful  for stimulating
discussions with Dr R Evans of BAE-Systems (Warton).

\vfill\eject

\appendix

\section{Definitions and Notation}
\label{sec:defs}

Throughout this article the geometry of spacetime is described in
terms of a  smooth Lorentzian metric tensor field and its
associated Levi-Civita connection. The connection 1-forms
$\omega^a{}_b$ associated with any such connection $\del$  are
defined with respect to any local basis of vector fields $\{X_a\}$
where $a=0,1,2,3$ by
\begin{equation}
  \del_{X_a}X_b=\omega^c{}_b(X_a)\, X_c.
\end{equation}


If $T$ is a smooth tensor field on spacetime then in terms of this
connection  the covariant differential $\del$ of $T$ is defined by
\begin{equation}
  (\del T)(X,-,...,-) = (\del_X T),
\end{equation}
and if the first argument of $T$ is contravariant, the divergence
of $T$ is
\begin{equation}
  \div T = (\del_{X_a} T)(X^a,-,...,-) ,
\end{equation}
where indices are raised with the components of the metric tensor
in the usual way. Similarly with $\{e^a\}$ dual to ${\{X_a\}}$
(i.e. $e^a(X_b)=\delta^a_b$)
\begin{equation}
  \div S = (\del_{X_a} S)(e^a,-,...,-) ,
\end{equation}
for all tensors $S$ whose first argument is covariant.
 For any degree 2 covariant tensor $T$ the tensor $\Sym T$
is defined by $\Sym T(X,Y)=\tfrac{1}{2}(T(X,Y)+T(Y,X))$.

In terms of the exterior derivative $\rmd$ on smooth differential
$p$-forms on spacetime,  the {\em covariant exterior derivative}
is defined on mixed basis indexed $p$-forms by
\begin{equation}
\begin{aligned}
  \rmD S^{a...b}{}_{c...d} = & \rmd S^{a...b}{}_{c...d} + \omega^a{}_s
  \wedge S^{s...b}{}_{c...d} + ... + \omega^b{}_s \wedge
  S^{a...s}{}_{c...d} \\
  & - \omega^s{}_c \wedge S^{a...b}{}_{s...d} -
  ... - \omega^s{}_d \wedge S^{a...b}{}_{c...s} .
\end{aligned}
\end{equation}
The curvature operator $\bfR$ for the connection $\del$ is
\begin{equation}
   \bfR_{X\:Y} = [\del_X,\del_Y] - \del_{[X,Y]}
\end{equation}
for all vector fields $X,Y$, and the (Riemann) curvature tensor
$R$ defined by
\begin{equation}
  R(X,Y,Z,\alpha) = \alpha(\bfR_{X\:Y} Z).
\end{equation}
Then in dual local bases $\{X_a\}$ and  $\{e^a\}$
the curvature 2-forms $R^a{}_b$ are defined by
\begin{equation}
  R = 2 R^a{}_b \tens e^b \tens X_a.
\end{equation}
 Successive contractions (where $\rmi_X$ is the interior operator
   with respect to the vector $X$) give the
Ricci 1-forms
\begin{equation}
  P_b = \rmi_{X_a} R^a{}_b ,
\end{equation}
and the curvature scalar $\cR$
\begin{equation}
  \cR = \rmi_{X^a} P_a .
\end{equation}

With $g$ a metric tensor on vectors and $G$ the induced   metric
tensor on 1-forms,  the metric-dual of a vector $X$ is given by
$\flatten{X} = g(X,-)$ and that  of and a 1-form $\alpha$ by
$\sharpen{\alpha} = G(\alpha,-)$.

If $U$ is a unit-normalized vector field ($g(U,U) = \lambda$ where
$\lambda = \pm 1$), then the projection operator on contravariant
tensor fields is defined as
\begin{subequations}
  \begin{equation}
     \bsPi_U = \bfone - \lambda U \tens \flatten{U} .
  \end{equation}
  By abuse of notation the same symbol is used to denote the projector on  covariant tensors
  \begin{equation}
     \bsPi_U = \bfone - \lambda \flatten{U} \tens U ,
  \end{equation}
  and differential  $p$-forms
  \begin{equation}
     \bsPi_U = \bfone - \lambda \flatten{U} \wedge \rmi_U ,
  \end{equation}
since the domain should be clear from the context. The map
$\bsPi_U$ is a tensor homomorphism
  \begin{equation}
     \bsPi_U (\ttt{\alpha}{\beta}) = \ttt[\bsPi_U]{\alpha}{\beta} ,
  \end{equation}
\end{subequations}
for all $\alpha, \beta, ...$ and has the following properties:
\begin{subequations}
  \begin{equation}
     \bsPi_U \bsPi_U = \bsPi_U ,
  \end{equation}
  \begin{equation}
     \bsPi_U \rmi_U = \rmi_U ,
  \end{equation}
  \begin{equation}
     \rmi_U \bsPi_U = 0 .
  \end{equation}
\end{subequations}
The projection tensor $\bsPi_V$ is used to define the {\em spatial
metric} on spacetime associated with the timelike vector field
$V$:
\begin{equation}
  \bfg_V = \Pi_V g .
\end{equation}
One  can write $\del \flatten{V}$ in terms of $\bfg_V$ and
$\bsPi_V$ as follows:
\begin{equation}
  \del \flatten{V} = \bssigma_V + \bsOmega_V
  + \tfrac{1}{3} \Theta_V \bfg_V
  - \flatten{V} \tens \flatten{\bscA}_V
\end{equation}
where
\begin{subequations}
  \begin{equation}
     \bssigma_V = \Sym \Pi_V \del \flatten{V} - \tfrac{1}{3} \Theta_V \bfg_V ,
  \end{equation}
is the {\em shear} of $V$,
  \begin{equation}
     \Theta_V = \div V ,
  \end{equation}
is the {\em expansion} of $V$,
  \begin{equation}
     \bsOmega_V = \Pi_V \rmd \flatten{V} ,
  \end{equation}
is the {\em vorticity} of $V$ and
  \begin{equation}
     \bscA_V = \del_V V
  \end{equation}
is the {\em acceleration} of $V$.
\end{subequations}

A vector field $\xi$ is said to be {\em parallel} (\wrt $\del$) if
\begin{equation}
  \del \xi = 0 ,
\end{equation}
in which case  $\del \flatten{\xi} = 0$, so  $\xi$ has vanishing
shear, vorticity, expansion and acceleration. Furthermore
\begin{equation}
  \Sym \del \flatten{\xi} = 0 ,
\end{equation}
or equivalently
\begin{equation}
  \cL_\xi g = 0
\end{equation}
in terms of the Lie derivative. Thus, if $\xi$ is parallel then it
is also a Killing vector and
\begin{equation}
  \label{eq:dell}
  \del_\xi = \cL_\xi .
\end{equation}

For any tensor field $T$ taking at least one covariant argument
let $T_\xi$ be defined by contraction such that
\begin{equation}
  T_\xi(-,\hdots,-) = T(-,\hdots,-,\xi,-,\hdots,-).
\end{equation}
If $\xi$ is parallel it follows that
\begin{equation}
  (\del_X T_\xi)(-,\hdots ,-)
  = (\del_X T)(-,\hdots ,-,\xi,-,\hdots ,-) \label{eq:divproperty}
\end{equation}
for all vector fields $X$.
When acting on $p$-forms this is simply the rule:
\begin{equation}
  \rmi_\xi \del_X = \del_X \rmi_\xi .
\end{equation}

The metric tensor $g$ gives rise to a canonical {\em volume}
4-form $\ast 1$ on spacetime and an associated Hodge map $\ast$ on
$p$-forms. In terms of a local $g-$orthonormal local basis of
1-forms $\{e^a\}$ one may write $\ast 1=e^0\wedge e^1\wedge
e^2\wedge e^3$. A volume 3-form $\# 1$ associated with the unit
timelike vector field $V$ is given in terms of  $\ast 1$  by
\begin{equation}
  \ast 1 = \flatten{V} \wedge \# 1
\end{equation}
with $\rmi_V\# 1=0$. Hence
\begin{equation}
  \# 1 = - \ast \flatten{V}
\end{equation}
and  $\# 1$ induces a spatial Hodge map, $\#$, on the image of
forms under $\bsPi_V$. With these operations defined one can
decompose the Hodge map of any spacetime form and its exterior
derivative into spatial forms. Thus  the Hodge dual of {\em any}
$p$-form $\omega$ on spacetime may always be written
\begin{equation}
  \label{eq:hodgesplit}
  \ast \omega = \Bigl\{ \# \rmi_V \omega \Bigr\}
  + \flatten{V} \wedge \Bigl\{ \# (\bsPi_V \omega)^\eta \Bigr\}
\end{equation}
where $\alpha^\eta=(-1)^p\alpha$ for any $p$-form $\alpha$ and
each of the  terms in brackets is annihilated by $V$  (i.e.
$\rmi_V \Bigl\{ \quad \Bigr\}=0$). A projected   Lie derivative
\wrt $V$ is defined as
\begin{equation}
  \bscL_V = \bsPi_V \cL_V \bsPi_V
\end{equation}
and for any p-form  $\omega$ one can write
\begin{equation}
  \cL_V \omega = \Bigl\{ \bscL_V \omega
  - \flatten{\bscA}_V \wedge \rmi_V \omega \Bigr\}
  - \flatten{V} \wedge \Bigl\{ \bscL_V \rmi_V \omega \Bigr\}
\end{equation}
where each of the  terms in brackets is annihilated by $V$.

In terms of the projected exterior derivative
\begin{equation}
  \bfd_V = \bsPi_V \rmd \bsPi_V ,
\end{equation}
and with $\bfD_V$ defined by
\begin{equation}
   \mathbf{D}_V = \mathbf{d}_V +
   \boldsymbol{\mathcal{A}}_V^\flat \wedge
\end{equation}
one may write
\begin{equation}
  \label{eq:dsplit}
  \rmd \omega = \Bigl\{ \bfd_V \omega
  - \bsOmega_V \wedge \rmi_V \omega \Bigr\}
  - \flatten{V} \wedge \Bigl\{ (\bscL_V \omega
  - \bfD_V \rmi_V \omega) \Bigr\}
\end{equation}
where each of the  terms in brackets is annihilated by $V$. These
formulae permit a local ``3+1'' decomposition of exterior
differential equations with respect to the general observer field
$V$ in a spacetime with an arbitrary Lorentzian metric and permit
one to identify  spatial fields parametrised by a local time
associated with $V$.

With the Faraday 2-form $F$ decomposed as
\begin{equation}
  F = \flatten{V} \wedge \bfe + \# \bfb
\end{equation}
where $\rmi_V \bfe = 0$, and $\rmi_V \bfb = 0$, it follows that
$\rmi_V F = - \bfe$ and $\bsPi_V F = \# \bfb$.  Similarly using
\eqn{hodgesplit}
\begin{equation}
  \ast F = \flatten{V} \wedge \bfb - \# \bfe ,
\end{equation}
so  $\rmi_V \ast F = - \bfb$ and $\bsPi_V F = \# \bfe$. Writing
\begin{equation}
  \cJ = \rho \flatten{V} + \bfj
\end{equation}
where $\rmi_V \bfj = 0$, and using \eqn{hodgesplit} it follows
that
\begin{equation}
  \ast \cJ = - \rho \# 1 - \flatten{V} \wedge \# \bfj .
\end{equation}

The co-derivative $\delta$ is defined on spacetime $p$-forms in
terms of the Hodge map $\ast$ and the exterior derivative  $\rmd$
by
\begin{equation}
  \delta = \ast^{-1} \rmd \ast \eta
\end{equation}
where $\eta \omega = \omega^\eta$ for any $p$-form $\omega$. The
field equations \eqn{maxcov1} and \eqn{maxcov2} can be written as
\begin{subequations}
  \begin{equation}
     \rmd F = 0 ,
  \end{equation}
  \begin{equation}
     \rmd \ast F = \ast \cJ .
  \end{equation}
\end{subequations}
Using \eqn{dsplit}  the  Maxwell equations given in \eqn{maxall}
follow immediately.

With the  definitions:
\begin{samepage}
  \begin{subequations}
    \begin{equation}
      \mathrm{div} \, \bfX = \sharpen{\# \bfd_V \# \flatten{\bfX}} ,
    \end{equation}
    \begin{equation}
      \mathrm{curl} \, \bfX = \sharpen{\# \bfd_V \flatten{\bfX}} ,
    \end{equation}
    \begin{equation}
      \mathrm{grad} \, \Phi = \sharpen{\bfd_V \Phi} ,
    \end{equation}
    \begin{equation}
      \bfX \times \bfY = \sharpen{\flatten{\bfX} \wedge \flatten{\bfY}} ,
    \end{equation}
    \begin{equation}
      \bfX \cdot \bfY = \bfg_V(\bfX, \bfY) =  \# (\flatten{\bfX} \wedge \# \flatten{\bfY})
    \end{equation}
  \end{subequations}
\end{samepage}
where $\bfX$ and $\bfY$ are spacelike (\wrt V) vectors, and $\Phi$
is a $0$-form on spacetime, exterior equations can be transcribed
to Euclidean vector notation.

The Laplacian operator $\Lap$ associated with $g$ and $\del$ is
defined by
\begin{equation}
  \Lap = \div \del ,
\end{equation}
and the associated Laplace-Beltrami operator $\Delta$ on $p$-forms
is
\begin{equation}
  \Delta = -(\delta \rmd + \rmd \delta) .
\end{equation}
It can be shown that, when acting on any $p$-form $\alpha$, $\Lap$
is related to the Laplace-Beltrami operator $\Delta$ by
\begin{equation}
  \Delta \alpha = \Lap \alpha + e^a \wedge \rmi_{X^b} \bfR_{X_a\:X_b} \alpha
  .
\end{equation}
Thus if $\beta$ is a 1-form
\begin{equation}
  \Delta \beta = \Lap \beta - \Ric(\sharpen{\beta},-)
\end{equation}
and for any 0-form $f$
\begin{equation}
  \Delta f = \Lap f .
\end{equation}
In a spacetime with a flat metric
\begin{equation}
  \Delta = \Lap
\end{equation}
for {\em all} $p$-forms.

\vfill\eject

\section{The Perturbed Connection and Gravito-Electromagnetic Fields}
\label{sec:connex}

With the $(3,0)$ tensor field $\flatten{\gamma}$ defined by
\begin{equation}
  \flatten{\gamma}(X,Y,Z) = \gamma(X,Y,\flatten{Z}) ,
\end{equation}
using \eqn{pertconnex} it follows that
\begin{equation}
  \label{eq:gammaxi}
  \flatten{\gamma}(\xi,-,-) = \tfrac{1}{2} \del_\xi h + \Alt (\del h)(-,\xi,-) ,
\end{equation}
where $\xi$ defines a \gemic frame (unit timelike  and parallel,
to zero order in $\epsilon$), and for any type (2,0) tensor $T$,
$\Alt T$ is defined by $\Alt T(X,Y)=\tfrac{1}{2}(T(X,Y)-T(Y,X))$.

To relate  expression \eqn{gammaxi} to \gemic fields one may
proceed as follows. By trace-reversing $\psi$ as given in
\eqn{psisplit}, $h$ can be written as
\begin{equation}
  \label{eq:hsplit}
  h = - \psi_\xi \tens \flatten{\xi} - \flatten{\xi} \tens \psi_\xi
  - \bsSigma_\xi + \tfrac{1}{2} \Tr(\bsSigma_\xi) g ,
\end{equation}
and acting  with $\del_\xi$ gives
\begin{equation}
  \del_\xi h = - \del_\xi \psi_\xi \tens \flatten{\xi} - \flatten{\xi} \tens \del_\xi \psi_\xi
  - \del_\xi \bsSigma_\xi + \tfrac{1}{2} \xi \Tr(\bsSigma_\xi) g .
\end{equation}

Since $\del \xi$ is first order in $\epsilon$ it follows that
\begin{equation}
  (\del h)(-,\xi,-) = \del(h(\xi,-)) ,
\end{equation}
and contracting \eqn{hsplit} with $\xi$ gives
\begin{equation}
  \label{eq:hxipsixi}
  h(\xi,-) = \psi_\xi - (\phi_\xi - \tfrac{1}{2}\Tr(\bsSigma_\xi))
  \flatten{\xi}.
\end{equation}
Hence
\begin{equation}
  (\del h)(-,\xi,-) = \del \psi_\xi - \rmd (\phi_\xi - \tfrac{1}{2}\Tr(\bsSigma_\xi)) \tens \flatten{\xi} .
\end{equation}
Antisymmetrizing this yields
\begin{equation}
  \Alt (\del h)(-,\xi,-) = \cF_\xi + \flatten{\xi} \wedge \bfd_\xi (\phi_\xi - \tfrac{1}{2}\Tr(\bsSigma_\xi)) ,
\end{equation}
where $\cF_\xi = \rmd \psi_\xi$.

Equation \eqn{gammaxi} can now be written as
\begin{equation}
  \label{eq:gammaxifinal}
  \begin{aligned}
  \flatten{\gamma}(\xi,-,-) = & - (\del_\xi \psi_\xi \tens \flatten{\xi} + \flatten{\xi} \tens \del_\xi \psi_\xi
  + \del_\xi \bsSigma_\xi) \\
  & + \tfrac{1}{4} \xi \Tr(\bsSigma_\xi) g
  + \cF_\xi + \flatten{\xi} \wedge \bfd_\xi (\phi_\xi - \tfrac{1}{2}\Tr(\bsSigma_\xi)) ,
  \end{aligned}
\end{equation}
and contracting on $\xi$ gives
\begin{equation}
  \flatten{\gamma}(\xi,\xi,-) = - \xi (\phi_\xi - \tfrac{1}{4} \Tr(\bsSigma_\xi)) \flatten{\xi}
  + \rmi_\xi \cF_\xi + \tfrac{1}{4} \bfd_\xi \Tr(\bsSigma_\xi) .
\end{equation}

If $\bfX$ is orthogonal to $\xi$, $g(\xi,\bfX) = 0$, and $\del
\bfX$ is first order or higher in $\epsilon$, then
$\flatten{\gamma}(\xi,\bfX,-)$ can be rewritten as
\begin{equation}
  \begin{aligned}
    \flatten{\gamma}(\xi,\bfX,-) = & - \xi(\psi_\xi(\bfX)) \flatten{\xi}
    + \xi (\tfrac{1}{4} \Tr(\bsSigma_\xi) + \bsSigma_\xi(\bfX,\bfX)) \flatten{\bfX} \\
    & + \rmi_\bfX (\rmi_\xi \cF_\xi + \tfrac{1}{4} \bfd_\xi \Tr(\bsSigma_\xi)) \xi
    + \tfrac{1}{2} \bsPi_\xi \rmi_\bfX \cF_\xi + \bsPi_\bfX \del_\xi \bsSigma_\xi(\bfX,-) .
  \end{aligned}
\end{equation}


Let $\xi$ be a vector field such that $\del \xi = 0$ and
$g(\xi,\xi) = -1$ (to at least first order in $\epsilon$). Using
\eqn{connex} and adding hats to quantities defined with respect to
the physical metric $\ghat$ it follows that
\begin{equation}
  \hat{\del} \xi = \gamma(\xi,-,-) ,
\end{equation}
or with $X^{\hat{\flat}} = \hat{G}(X,-)$, that:
\begin{equation}
  \hat{\del} \xi^{\hat{\flat}} = \flatten{\gamma}(\xi,-,-) ,
\end{equation}
where $\flatten{\gamma}(X,Y,Z) = \gamma(X,Y,\flatten{Z})$.
Antisymmetrizing and using \eqn{gammaxifinal} yields
\begin{equation}
  \label{eq:antisymdxi}
  \rmd \xi^{\hat{\flat}} = \cF_\xi + \flatten{\xi} \wedge \bfd_\xi (\phi_\xi - \tfrac{1}{2} \Tr(\bsSigma_\xi)) .
\end{equation}
The left hand side can be contracted on $\xi$ and rewritten as
\begin{equation}
  \label{eq:accelhat}
  \rmi_\xi \rmd \xi^{\hat{\flat}} = \hat{\bscA\,\,\,}^{\hat{\flat}}_\xi - \tfrac{1}{2} \rmd (h(\xi,\xi)) ,
\end{equation}
where $\hat{\bscA\,\,\,}_\xi = \hat{\del}_\xi \xi$. Contracting
\eqn{hxipsixi} with $\xi$ and using \eqn{antisymdxi} it follows
that
\begin{equation}
  \label{eq:accelxi}
  \hat{\bscA\,\,\,}^{\hat{\flat}}_\xi = - \bscE_\xi + \tfrac{1}{4} \bfd_\xi \Tr(\bsSigma_\xi)) + \tfrac{1}{4} \xi \Tr(\bsSigma_\xi) \flatten{\xi} .
\end{equation}
Defining
\begin{equation}
  \hat{\bsOmega}_\xi = \hat{\bsPi}_\xi \rmd \xi^{\hat{\flat}} ,
\end{equation}
and projecting \eqn{antisymdxi} gives
\begin{equation}
  \hat{\bsOmega}_\xi = \# \bscB_\xi .
\end{equation}

In the \gemic limit and with $\xi \phi_\xi = 0$ \eqn{accelxi}
simplifies to
\begin{equation}
  \hat{\bscA\,\,\,}^{\hat{\flat}}_\xi = - \tfrac{1}{4} \bscE_\xi .
\end{equation}

Thus the \gemic fields can be interpreted in terms of the
vorticity and acceleration of the vector field $\xi$ with respect
to $\ghat$.

\vfill\eject

\section{Transformation Laws For Gravito-Electromagnetic Fields Under a Change of Frame}
\label{sec:trans}

Let $\zeta$ and $\xi$ define two \gemic frames (as defined in
section \sect{gauge}) such that
\begin{equation}
  \zeta = \xi + \bfv ,
\end{equation}
where $g(\xi,\bfv) = 0$ and $v^2 = g(\bfv,\bfv) \ll 1$.

Since $\psi$ is independent of any frame of reference it may be
expanded as
\begin{subequations}
  \begin{equation}
    \psi = \phi_\xi g - \psi_\xi \tens \flatten{\xi} - \flatten{\xi} \tens \psi_\xi - \bsSigma_\xi ,
  \end{equation}
in terms of $\zeta$ or as
  \begin{equation}
        \psi = \phi_\zeta g - \psi_\zeta \tens \flatten{\zeta} -
        \flatten{\zeta} \tens \psi_\zeta - \bsSigma_\zeta
  \end{equation}
\end{subequations}
in terms of $\xi$.
 Equating these two expressions
yields
\begin{subequations}
  \begin{equation}
    \psi_\zeta = \psi_\xi - \psi_\xi (\bfv) \flatten{\xi} + \left\{ \phi_\xi \flatten{\bfv} - \bsSigma_\xi(\bfv,-) \right\} ,
  \end{equation}
  \begin{equation}
    \begin{aligned}
      \bsSigma_\zeta = & \; \bsSigma_\xi + 2 \psi_\xi (\bfv) \bfg_\xi - \psi_\xi \tens \flatten{\bfv} - \flatten{\bfv} \tens \psi_\xi \\
      & - \left\{ \phi_\xi \flatten{\bfv} - \bsSigma_\xi(\bfv,-) \right\} \tens \flatten{\xi} - \flatten{\xi} \tens \left\{ \phi_\xi \flatten{\bfv} - \bsSigma_\xi(\bfv,-) \right\} ,
    \end{aligned}
  \end{equation}
\end{subequations}
Thus if $\bfv$ is of order $\epsilon$ the fields $\psi_\zeta$ and
$\bsSigma_\zeta$ are invariant under a change of \gemic frame  to
first order in $\epsilon$.

If the \gemic limit is satisfied in the $\xi$ frame,
 the terms in braces vanish and the above simplify to
\begin{subequations}
  \begin{equation}
    \psi_\zeta = \psi_\xi - \psi_\xi (\bfv) \flatten{\xi} ,
  \end{equation}
  \begin{equation}
    \bsSigma_\zeta = \{ \phi_\xi + 2 \psi_\xi (\bfv) \} \bfg_\xi -
     \psi_\xi \tens \flatten{\bfv} - \flatten{\bfv} \tens \psi_\xi .
  \end{equation}
\end{subequations}

If ${\bfv}$ is of order $\epsilon$ then  the \gemic limit is
preserved and the above fields remain invariant.

\vfill\eject


\section{An Alternative Analogy Between Einstein's Equations and
Electromagnetism Based on Properties of the Conformal Tensor}
\label{sec:conf}

%
%

As  mentioned in the introduction there exist alternative
analogies between Einstein's theory of gravitation and Maxwell's
electromagnetic equations that do not necessarily require a
perturbative approach. One such analogy is summarised here in
terms of the conformal tensor on spacetime since it highlights the
differences accorded to gravito-electromagnetism by different
authors \cite{maartens}, \cite{ehlers}, \cite{Bonnor1},
\cite{Bonnor2},
\cite{Ellis1}, \cite{Ellis2}. 

The fourth order (Weyl) conformal tensor  may  be written as
\begin{equation}
  \label{eq:conftens}
  C = - 2 C_{ab} \tens (e^a \wedge e^b)
\end{equation}
where  (in any local cobasis $\{e^a\}$),  the conformal 2-forms
$C_{ab}$ are defined  in terms of the curvature 2-forms $R_{ab}$,
the Ricci 1-forms $P_a$, \cite{rwt}  and the curvature scalar
$\cR$:
\begin{equation}
  C_{ab} = R_{ab} - \frac{1}{2}(P_a \wedge e_b - P_b \wedge
  e_a) + \frac{1}{6} \cR e_a \wedge e_b .
\end{equation}
The covariant exterior derivative of the conformal 2-forms is
\begin{equation}
  \rmD C_{ab} = - \frac{1}{2}(Y_a \wedge e_b - Y_b \wedge e_a) ,
\end{equation}
where the 2-forms $Y_a$ are defined by
\begin{equation}
  Y_a = \rmD P_a - \frac{1}{6} \rmd \cR \wedge e_a .
\end{equation}
     If the geometry of spacetime is determined by Einstein's
     equations then these forms may be related to
the stress energy-momentum tensor $\cT$ by introducing the 3-forms
$\tau_a$ such that
\begin{equation}
  \cT = (\ast^{-1} \tau_a) \tens e^a .
\end{equation}
Since the connection is torsion-free:
\begin{equation}
  Y_a = \rmD (P_a - \frac{1}{6} \cR \wedge e_a) .
\end{equation}
It follows immediately from Einstein's equations written in terms
of the Ricci forms $P_a$ \cite{rwt} that
\begin{equation}
  Y_a = \frac{\kappa}{2} \rmD \left(\ast^{-1} \tau_a - \tfrac{1}{3}( \rmi_{X^c} \ast^{-1} \tau_c) e_a\right ) ,
\end{equation}
where $\kappa = 8 \pi G$.
%
%

In terms of the {\em covariant Lie derivative} $\rmL_{X}$:
\begin{equation}
  \rmL_{X} = \rmD \rmi_{X} + \rmi_{X} \rmD ,
\end{equation}
one may show that
\begin{equation}
  \label{eq:confmax}
  \rmL_{X^a} C_{ab} = \frac{1}{2} Y_b
\end{equation}
and
\begin{equation}
  \label{eq:confcons}
  \rmL_{X^a} Y_a = 0 .
\end{equation}
These relations may be compared with the  equations for the
Faraday 2-form $F$:
\begin{equation}
  \label{eq:max}
  \delta F = \cJ ,
\end{equation}
from which there follows the conservation of electric current,
\begin{equation}
  \label{eq:cons}
  \delta \cJ = 0 .
\end{equation}
In any local  cobase $F$ can be written in terms of its components
as
\begin{equation}
  F = F_{ab} e^a \wedge e^b ,
\end{equation}
(compare  with \eqn{conftens}) and $\cJ$ as
\begin{equation}
  \cJ = \cJ_a e^a .
\end{equation}
In terms of the {\em vector-valued $0$-forms} $F_{ab}$ and $\cJ_a$
\eqn{max} and \eqn{cons}  can be rewritten as
\begin{equation}
  \rmL_{X^a} F_{ab} = - \frac{1}{2} \cJ_b ,
\end{equation}
\begin{equation}
  \rmL_{X^a} \cJ_a = 0
\end{equation}
which are analogous to \eqn{confmax} and \eqn{confcons}.

The use of local coframes is not mandatory to see this
correspondence.  A purely tensorial formulation follows by writing
\begin{equation}
  \div C = - Y ,
\end{equation}
where
\begin{equation}
  Y = e^a \tens Y_a .
\end{equation}
This is the analogue of the Maxwell equation
\begin{equation}
  \div F = - \frac{1}{2} \cJ .
\end{equation}
Similarly one may  write
\begin{equation}
  \div Y = 0 ,
\end{equation}
to compare it with the conservation of electric current written in
the form
\begin{equation}
  \div \cJ = 0 .
\end{equation}









\bibliographystyle{amsplain}
\bibliography{sjcrefs}


\end{document}